\documentclass{article}

\usepackage{PRIMEarxiv}

\usepackage[utf8]{inputenc} 
\usepackage[T1]{fontenc}    
\usepackage{hyperref}       
\usepackage{url}            
\usepackage{booktabs}       
\usepackage{amsfonts}       
\usepackage{nicefrac}       
\usepackage{microtype}      
\usepackage{lipsum}
\usepackage{fancyhdr}       
\usepackage{graphicx}       
\graphicspath{{media/}}     

\usepackage{subcaption}
\usepackage{amsmath}
\usepackage{siunitx}
\newcommand{\mI}{\mathbf{I}}

\newcommand{\vf}{\mathbf{f}}

\newcommand{\vu}{\mathbf{u}}
\newcommand{\vv}{\mathbf{v}}
\newcommand{\vx}{\mathbf{x}}
\newcommand{\vy}{\mathbf{y}}
\newcommand{\diff}{\mathrm{d}}
\usepackage{calc}
\usepackage{booktabs}
\usepackage{xcolor}
\definecolor{clrVibr}{RGB}{27, 158, 119}
\definecolor{clrOneLow}{RGB}{217, 95, 2}
\definecolor{clrTwoLow}{RGB}{117, 112, 179}
\definecolor{clrTwoHigh}{RGB}{231, 41, 138}
\usepackage{wasysym}
\newcommand{\makecircle}[1]{\textrm{\textcolor{#1}{\newmoon}}}%
\newcommand{\makecircleem}[1]{\emph{\textcolor{#1}{\newmoon}}}%

\title{Vocal Fold Reconstruction from Optical Velocity and
Displacement Measurements
}

\author{
  Daniel Zieger \\
  Department of Computer Science \\
  Friedrich-Alexander-Universit{\"a}t Erlangen-N{\"u}rnberg \\
  Erlangen, Germany\\
  \texttt{daniel.zieger@fau.de} \\
  \And
  Christoph N{\"a}ger \\
  Institute of Fluid Mechanics \\
  Friedrich-Alexander-Universit{\"a}t Erlangen-N{\"u}rnberg \\
  Erlangen, Germany\\
  \texttt{christoph.naeger@fau.de} \\
  \And
  Stefan Becker \\
  Institute of Fluid Mechanics \\
  Friedrich-Alexander-Universit{\"a}t Erlangen-N{\"u}rnberg \\
  Erlangen, Germany\\
  \texttt{stefan.becker@fau.de} \\
  \And
  Tobias G{\"u}nther \\
  Department of Computer Science \\
  Friedrich-Alexander-Universit{\"a}t Erlangen-N{\"u}rnberg \\
  Erlangen, Germany\\
  \texttt{tobias.guenther@fau.de} \\
}

\begin{document}
\maketitle

\begin{abstract}
The three-dimensional reconstruction of vocal folds in medicine usually involves endoscopy and an approach to extract depth information like structured light or stereo matching of images.
The resulting mesh can accurately represent the superior area of the vocal folds, while new approaches also try to reconstruct the inferior area.
We propose a novel approach to extract the time-dependent 3D geometry of the vocal fold from optical measurements on both the superior and inferior side, requiring optical measurements only from the superior side. First, a time-dependent, tri-variate surface velocity vector field is reconstructed using a high-speed camera and a laser Doppler vibrometer in an experimental environment.
This vector field serves as target in an inverse finite-element simulation that optimizes the forces applied to a deformable vocal fold model such that the resulting movement after FEM simulation matches the velocity observations on the superior side.
The required forces for the finite element method simulation are treated as unknowns and are assembled using multiple scalar fields.
We use tensor products in B\'ezier Bernstein basis for our scalar fields to reduce the degrees of freedom for our optimization.
We use gradient descent to optimize the control points of the force field polynomials.
Our utilized error metric for gradient descent consists of two terms.
The first term is used to match the simulated velocities to the observed measurements, while the second term measures the silhouette difference between observation and simulation.
\end{abstract}


\section{Introduction}
\label{sec:introduction}
Human communication is primarily based on the voice. Consequently, the voice has a great impact on the daily life of a person, as the quality of life of an individual is significantly decreased if they suffer from voice impairments \cite{Merrill2013}. Physically speaking, the human voice is generated in a complex interplay of fluid flow, structural vibration, and acoustics. This process is called phonation. During phonation, the vocal folds are excited by a flow of air $\dot{V}$ from the lungs, causing them to vibrate. The vocal fold oscillation in return entails a modulation of this airflow, resulting in a pulsating jet flow in the vocal tract. This modulated jet flow is the main acoustic source in phonation, as discussed by Kaltenbacher et al.~\cite{Kaltenbacher2014}. The generated sound is filtered through the vocal tract and radiated from the mouth, resulting in the voice. 
Even though the vocal fold vibration is not the main direct sound source in phonation, it still plays a very important part by its interaction with the air flow. As it is difficult to perform extensive measurements of the glottal air flow in vivo due to a restricted accessibility, clinicians often rely on measurements of the vocal fold vibration for the diagnosis of speech impairments. Endoscopic two-dimensional imaging is thereby the most established method. This has been extended to high-speed videoendoscopy with frame rates up to 20 kHz in the past to be able to resolve the full oscillation cycle of the vocal folds in time \cite{Patel2014,Echternach2013}. However, the work by D{\"o}llinger and Berry~\cite{Doellinger2006a} showed that there is a non-negligible out-of-plane movement present in the vocal fold vibration, which is not captured with conventional two-dimensional endoscopy. To be able to obtain a three-dimensional vocal fold movement, a stereo triangulation technique has been developed. For this, often a glass prism is used to generate two different viewpoints with one endoscope, cf.~\cite{Doellinger2006a}. Also two independent optical systems that are displaced by a known distance have been used~\cite{Sommer2014}. To facilitate feature matching, marker points can be generated by projecting a structured light pattern onto the vocal fold surface~\cite{Luegmair2010,Patel2013,Luegmair2015,Semmler2016}. To help practitioners with diagnosing laryngeal disorders, Semmler et al.~\cite{Semmler2016} developed a method of reconstructing the superior vocal fold surface as a 3D-model. This has been further developed by Henningson et al.~\cite{Henningson2022}, who reconstructed the inferior vocal fold surface via B-spline interpolation, resulting in a full 3D vocal fold model based on the M5-model by Scherer et al.~\cite{Scherer2001}. The inferior vocal fold surface movement here is however not necessarily physically valid, as no physical constraints are used when applying the B-spline model. Nevertheless, this is still a useful representation for helping practitioners making a diagnosis. For a more detailed physical analysis, however, it would certainly be more advantageous to apply a more physics-based approach of vocal fold reconstruction, which is the contribution of this paper. 

One application, in which 3D vocal fold models are currently used is the simulation of the phonation process via computational fluid dynamics (CFD). In fully coupled fluid-structure interaction simulations, the vocal fold movement is directly simulated by coupling CFD with a lumped element model or finite element simulation. As an example for the use of lumped elements, Erath et al.~\cite{Erath2011} used a two mass model of the vocal folds in combination with boundary layer methods for the flow. Using lumped elements is a strong simplification of the vocal fold movement.
One example for coupling a finite element solver to a Navier-Stokes based flow solver is the work by Zheng et al.~\cite{Zheng2011}. They applied the immersed boundary method to analyse the aerodynamics of the phonation process. This approach can generate very accurate results, while at the same time a significant amount of computational resources is necessary to perform such kind of simulations. For this reason, one-way coupled fluid-structure simulations have emerged as a useful method for simulating human phonation. Here, the vocal fold movement is being prescribed and therefore not influenced by the surface pressure arising from the intraglottal flow field. With this, the computationally costly fluid-structure interaction simulation is reduced to a pure CFD-simulation. This method was applied first by Z{\"o}rner et al.~\cite{Zoerner2013} and was developed further in several publications with the goal of making forward-coupled CFD-simulations clinically applicable for voice treatment~\cite{Sadeghi2019JVoice,Schoder2020,Falk2021,Maurerlehner2021}. Furthermore, an extensive analysis of the aeroacoustic sources in phonation could be performed by Schoder et al.~\cite{Schoder2021}, applying the perturbed convective wave equation.
These simulations used a two-parameter model for the vocal fold movement, such that the vocal fold motion has a translational part in medial-lateral direction and a rotational part generating the typical convergent-divergent glottal gap. The velocities of opening and closing the gap were chosen to reproduce the glottal area waveform of an experimental model. However, as stated before, a physiologically realistic vocal fold movement features a significant out-of-plane movement~\cite{Doellinger2006a}, which is not included in these models. Therefore, one-way coupled fluid-structure simulations could benefit from an improved prescribed vocal fold movement model. 

Considering these two main applications, the present work introduces a procedure to reconstruct a physically valid three-dimensional vocal fold model from three-dimensional velocity measurements. The velocity data is obtained experimentally from a synthetic vocal fold model via high-speed-camera imaging and laser-Doppler-vibrometry. Nonlinear optimization is applied to fit the movement of a finite-element-model of the vocal folds to the measured velocity data in an optimal sense. Additionally, the glottal area shape is extracted from the high-speed-camera images and included in the optimization procedure. The result is a three-dimensional vocal fold model that can be used as an input for future CFD-simulations of the phonation process with prescribed vocal fold movement. 
Since a finite-element-model serves as shape prior, the optimization of a full three-dimensional model (inferior and superior side) can be performed from optical measurements of the superior side alone.
Thus, the same procedure can also be used in the medical diagnosis of voice impairments via videoendoscopy by applying it to the patient specific vocal fold movement.
In general, the proposed procedure can be applied to reproduce any three-dimensional displacement field of a solid body from three-dimensional velocity data while at the same time satisfying the underlying physical equations. 

\begin{figure}[t]%
	\centering%
    \begin{subfigure}{.47\linewidth}
    \includegraphics[width=\linewidth]{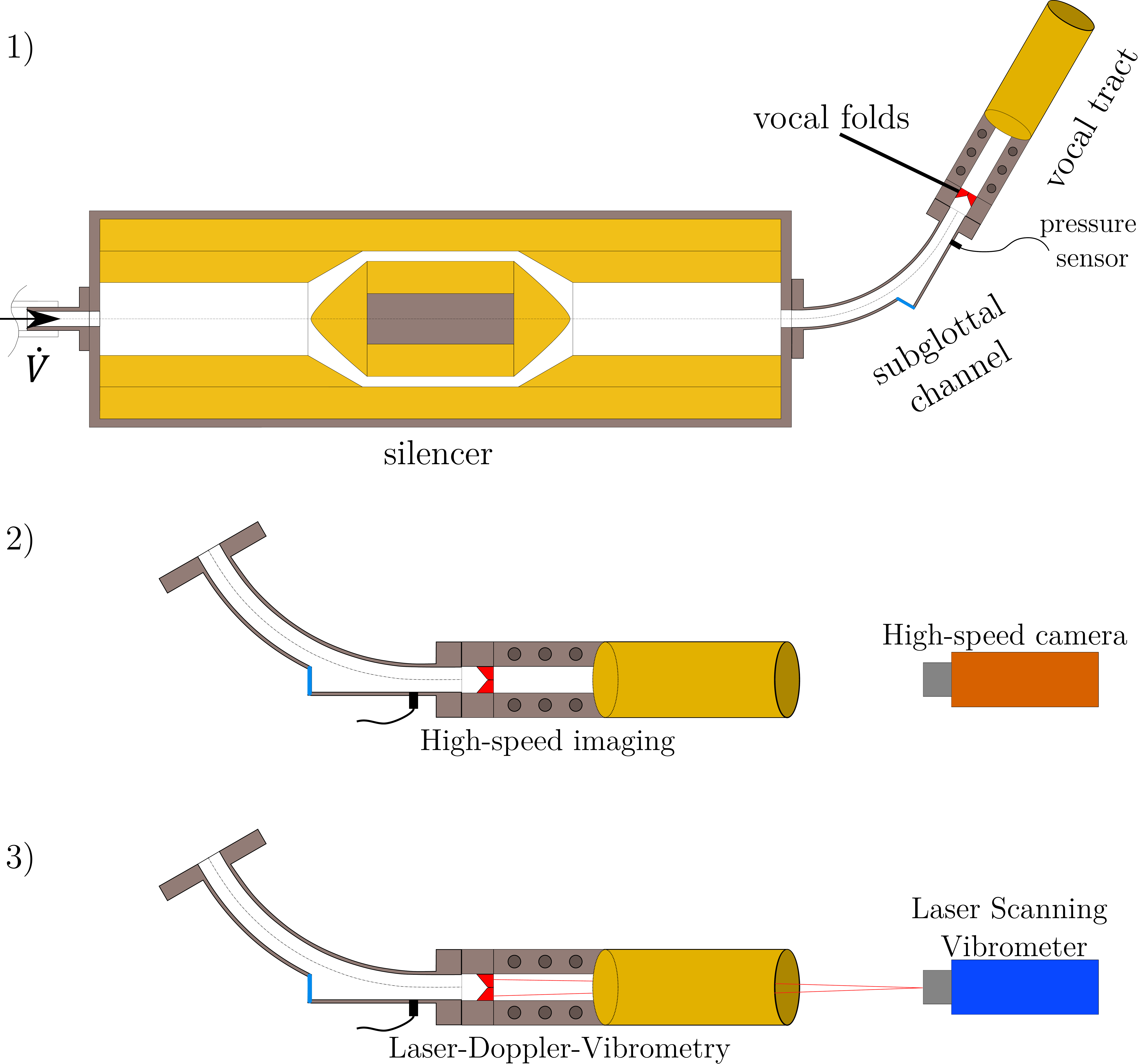}%
	\caption{
    Schematic illustration of the experimental setup. 
    }%
    \end{subfigure}%
    \hfill%
    \begin{subfigure}{.45\linewidth}%
    \raisebox{1.2cm}{%
    \includegraphics[width=\linewidth]{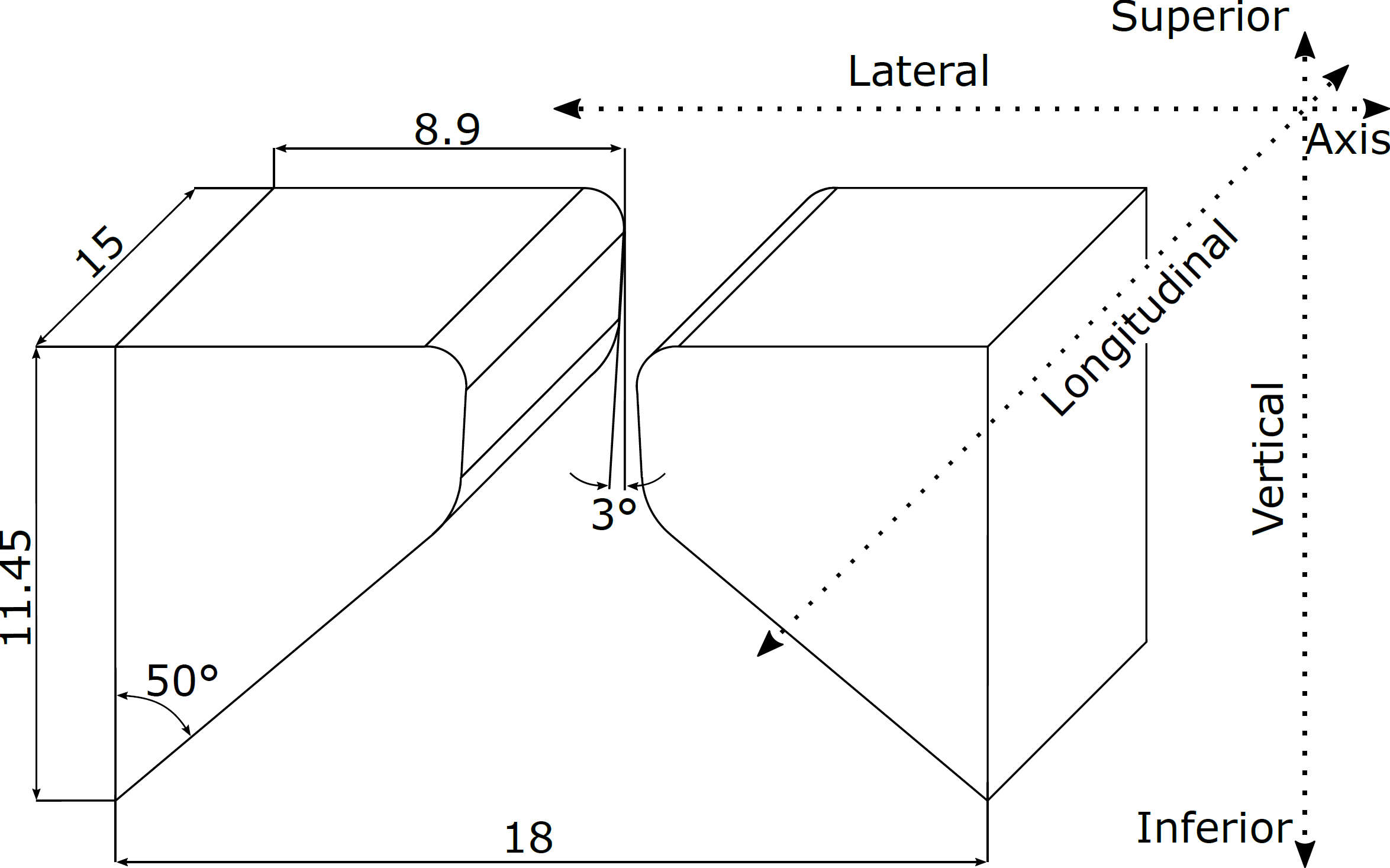}%
    }%
	\caption{
        Shape parameters of the vocal fold model.
    }%
    \end{subfigure}
    \caption{Illustration of the experimental setup. In (a), the images show 1) the labeled parts of the entire setup, 2) the components of the high-speed imaging, and 3) the components for the Laser-Doppler-Vibrometry. The vocal folds are highlighted in red. In (b), shape parameters of the used vocal folds model are shown with the units given in millimeters.}
	\label{fig:expset}%
\end{figure}%

\section{Experimental Setup}
\label{sec:expSet}

The experimental setup is a slightly modified version of the setup presented by Lodermeyer et al.~\cite{Lodermeyer2015} and is shown in Fig. \ref{fig:expset}. A mass flow generator containing a supercritical valve (see \cite{Durst2003}) is responsible for the flow generation through the test rig. The total mass flow enters a silencer that dampens acoustic fluctuations coming from the supply hose. After the silencer, a special curved subglottal channel is added to allow for optical access to the inferior side of the vocal folds. After the vocal folds, a uniform vocal tract with a length of 200\,mm is attached. The length is chosen such that the acoustic resonance frequencies of the vocal tract do not evoke acoustically driven vocal folds as proposed by Zhang et al.~\cite{Zhang2006}.
The material of the vocal folds is an elastomer with a Young's modulus of  $E = 4.4$\,kPa (\cite{Lodermeyer2015}) and a Poisson's ratio of $\nu = 0.499$ (\cite{Rupitsch2011}).
Two different measurement methods were chosen to obtain the three-dimensional velocities at the superior as well as at the inferior vocal fold surfaces. They were conducted sequentially and synchronized in a post processing step by making use of the simultaneously acquired subglottal pressure signal. The subglottal pressure signal was recorded using a Kulite XCQ-093 pressure sensor (Kulite Semiconductor Products Inc., Leonia, NJ, USA) that was located 50mm upstream of the glottal exit at a measurement frequency of 50\,kHz.
Planar high-speed imaging of the superior vocal fold surface was performed to obtain information of the vocal fold movement in the transversal plane.  The images were acquired by a Photron SA-X2 high speed camera (Photron, Tokyo, Japan) in combination with a AF-S DX NIKKOR 18-300\,mm 1:3.5-6.3G ED VR (Nikon, Tokyo, Japan) at a rate of 10,000 frames per second.
For the out-of-plane movement, Laser-Doppler-Vibrometry was applied, using a Polytec PSV-500 Scanning Vibrometer (Polytec, Waldbronn, Germany). A grid of $17 \times 22$ measurement points on the superior vocal fold surface was chosen, resulting in a total of 374 measurement points. The measurement frequency of the vibrometer was set to 50\,kHz.

\begin{figure*}[t]%
    \centering%
    \includegraphics[width=\linewidth]{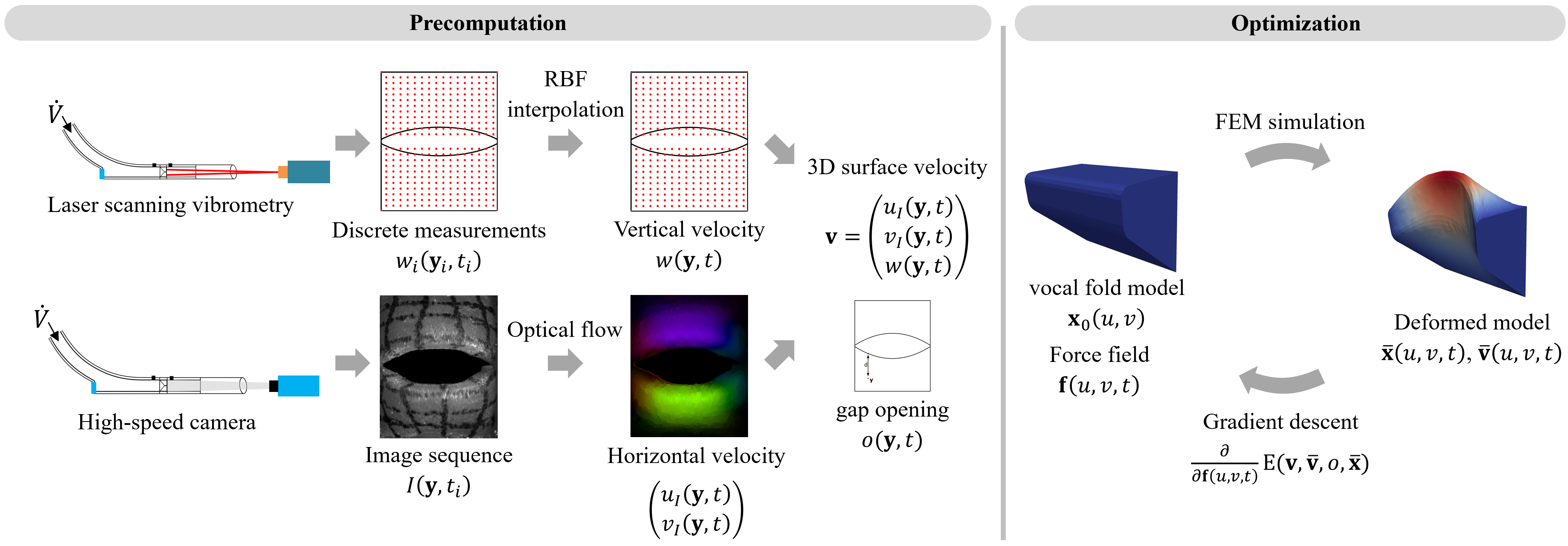}%
    \caption{Our experimental setup delivers high-speed camera images $I(\vy_i, t_i)$ and discrete vertical velocity measurements $w_i(\vy_i,t_i)$. In a precomputation, we reconstruct the horizontal components $(u_I(\vy,t), v_I(\vy,t))$ and the vertical component $w(\vy,t)$ of the surface velocity field $\vv$, as well as an unsigned distance field $\sigma(\vy,t)$ that measures the distance to the vocal fold opening. We use a finite element (FEM) simulation that takes an initial vocal fold model $\vx_0(u,v)$ and a surface force field $\vf(u,v,t)$ as input and calculates a deformed model with positions $\overline\vx(u,v,t)$ and velocities $\overline\vv(u,v,t)$. Using gradient descent, we optimize for the force field $\vf(u,v,t)$ such that the simulated vocal fold motion and the opening match the observations, i.e., the measurements that were processed in the precomputation.}%
    \label{fig:overview}%
\end{figure*}

\section{Vocal Fold Reconstruction}
\label{ch:meth}

\subsection{Overview}
Our computational vocal fold reconstruction pipeline is divided into two steps: (1) precomputation and (2) optimization.
A schematic illustration is provided in Fig.~\ref{fig:overview}.
In the precomputation, the discrete vibrometer measurements are interpolated in space and time, resulting in the vertical component of the vocal fold target surface velocity field.
Further, the horizontal component of the target surface velocity field is estimated using an optical flow algorithm from the high-speed camera footage.
In addition, a time-dependent unsigned distance field is computed that measures the distance to the silhouette of the opening, i.e., the gap.
The optimization takes a deformable vocal fold model, the reconstructed fields, and a force vector field as input. 
The latter is applied in a finite-element method simulation to the initial vocal fold model, which results in a deformed vocal fold model.
From the deformed model, the resulting surface vector field and the resulting opening are compared to the optical measurements that were precomputed.
Via gradient descent, the force field is gradually improved to reduce the distance between the deformed vocal fold model and the measurement data.
In the following sections, the individual steps are described formally in more detail.

\subsection{Reconstruction of Surface Vector Field from Optical Measurements}
In this section, we reconstruct a continuous, time-dependent velocity vector field $\vv(\vy,t) : \mathbb{R}^2\times \mathbb{R} \rightarrow \mathbb{R}^3$ that describes the screen-space motion of the observed vocal fold surface from the experimental setup described in Section \ref{sec:expSet}.
We combine this vector field later with a strong shape prior of the vocal folds using the finite-element method (FEM).

\begin{figure}[b]
    \centering
    \begin{subfigure}{.125\textwidth}
        \centering
        \scriptsize{$6.5\;\si{ms}$}\\
        \includegraphics[width=1\linewidth]{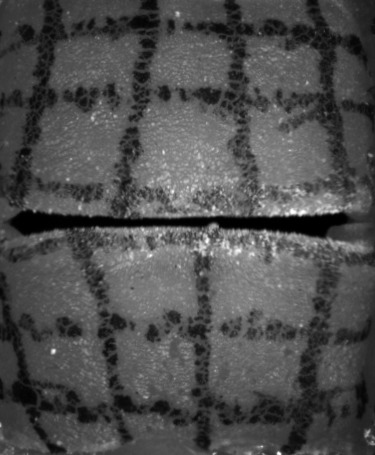}
    \end{subfigure}
    \begin{subfigure}{.125\textwidth}
        \centering
        \scriptsize{$6.6\;\si{ms}$}\\
        \includegraphics[width=1\linewidth]{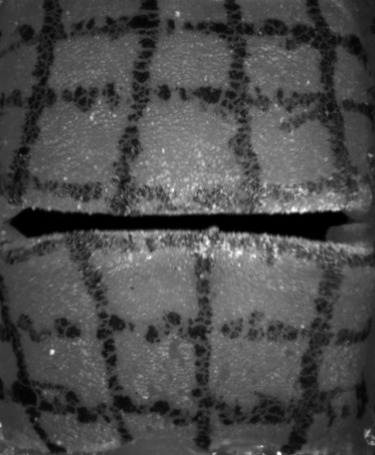}
    \end{subfigure}
    \begin{subfigure}{.125\textwidth}
        \centering
        \scriptsize{$6.7\;\si{ms}$}\\
        \includegraphics[width=1\linewidth]{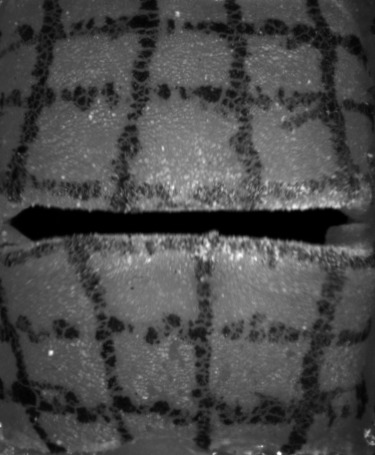}
    \end{subfigure}
    \begin{subfigure}{.125\textwidth}
        \centering
        \scriptsize{$6.8\;\si{ms}$}\\
        \includegraphics[width=1\linewidth]{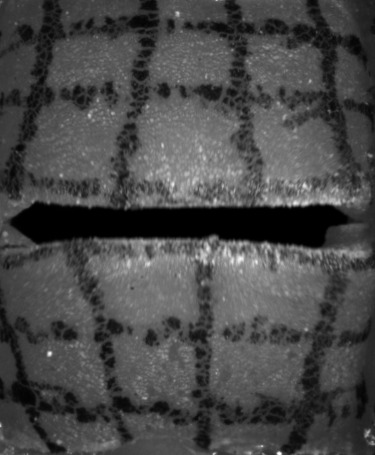}
    \end{subfigure}
    \begin{subfigure}{.125\textwidth}
        \centering
        \scriptsize{$6.9\;\si{ms}$}\\
        \includegraphics[width=1\linewidth]{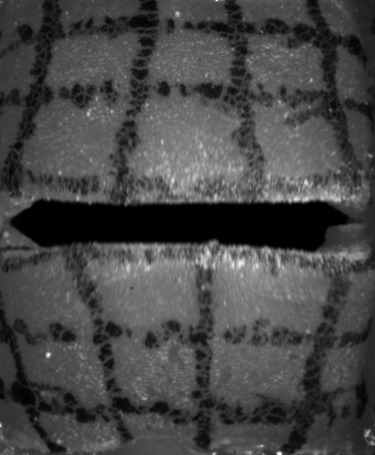}
    \end{subfigure}
    \begin{subfigure}{.125\textwidth}
        \centering
        \scriptsize{$7.0\;\si{ms}$}\\
        \includegraphics[width=1\linewidth]{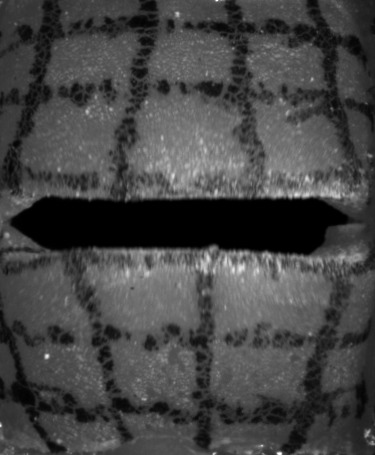}
    \end{subfigure}
    \begin{subfigure}{.125\textwidth}
        \centering
        \scriptsize{$7.1\;\si{ms}$}\\
        \includegraphics[width=1\linewidth]{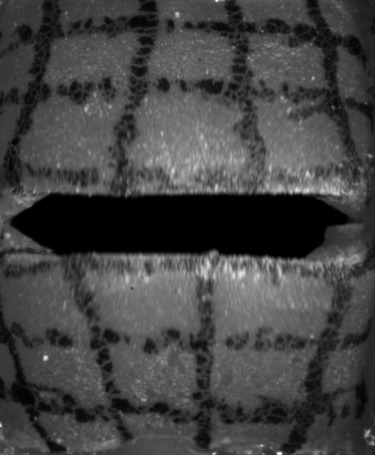}
    \end{subfigure}
    \begin{subfigure}{.125\textwidth}
        \centering
        \vspace{0.5em}
        \scriptsize{$7.2\;\si{ms}$}\\
        \includegraphics[width=1\linewidth]{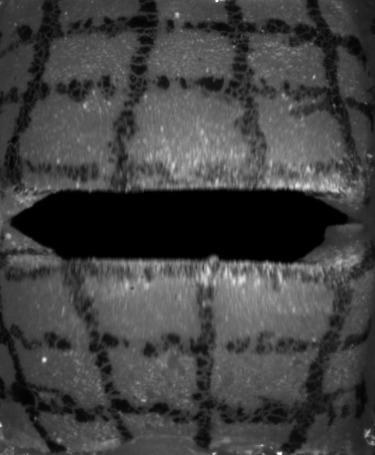}
    \end{subfigure}
    \begin{subfigure}{.125\textwidth}
        \centering
        \scriptsize{$7.3\;\si{ms}$}\\
        \includegraphics[width=1\linewidth]{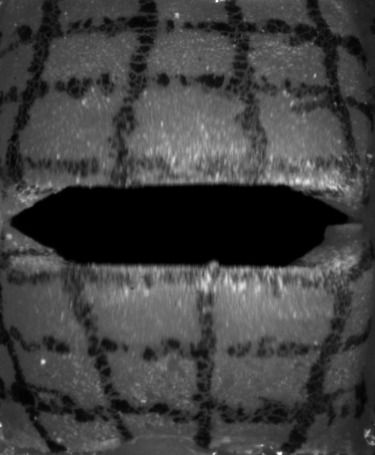}
    \end{subfigure}
    \begin{subfigure}{.125\textwidth}
        \centering
        \scriptsize{$7.4\;\si{ms}$}\\
        \includegraphics[width=1\linewidth]{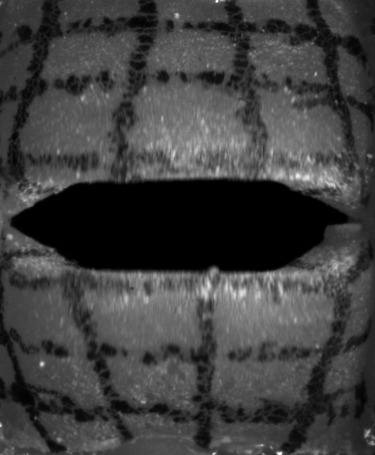}
    \end{subfigure}
    \begin{subfigure}{.125\textwidth}
        \centering
        \scriptsize{$7.5\;\si{ms}$}\\
        \includegraphics[width=1\linewidth]{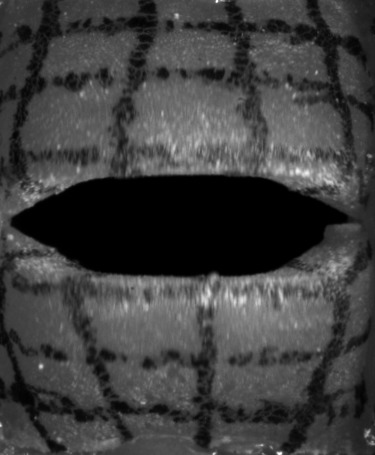}
    \end{subfigure}
    \begin{subfigure}{.125\textwidth}
        \centering
        \scriptsize{$7.6\;\si{ms}$}\\
        \includegraphics[width=1\linewidth]{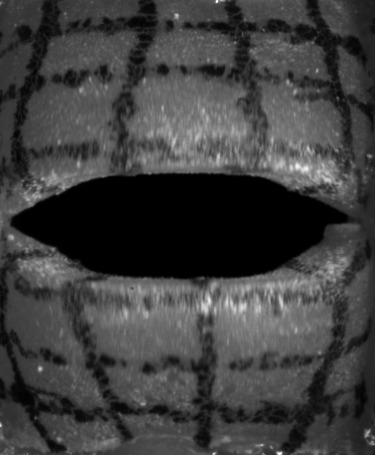}
    \end{subfigure}
    \begin{subfigure}{.125\textwidth}
        \centering
       \scriptsize{ $7.7\;\si{ms}$}\\
        \includegraphics[width=1\linewidth]{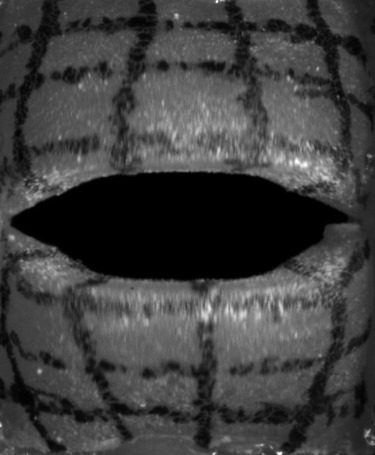}
    \end{subfigure}
    \begin{subfigure}{.125\textwidth}
        \centering
        \scriptsize{$7.8\;\si{ms}$}\\
        \includegraphics[width=1\linewidth]{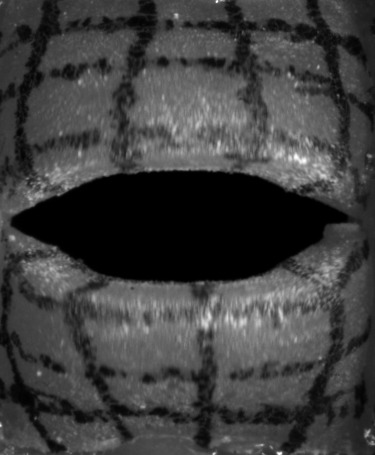}
    \end{subfigure}
    \caption{
    Time series of images of the superior surface of the vocal folds captured by a high-speed camera. 
    }
    \label{fig:vfhsc}
\end{figure}

\begin{figure}[t]
    \centering
    \begin{subfigure}{.33\textwidth}
        \centering
        \includegraphics[width=0.5\linewidth]{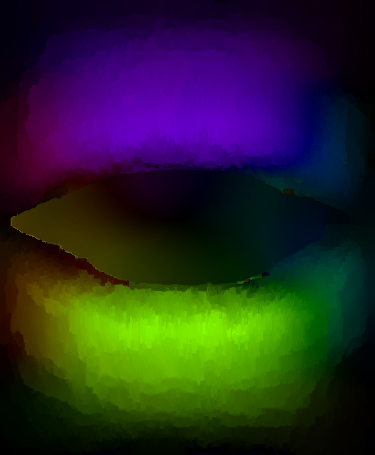}
    	\caption{Oversmoothed optical flow}
    	\label{fig:osof}
    \end{subfigure}
    \begin{subfigure}{.33\textwidth}
        \centering
        \includegraphics[width=0.5\linewidth]{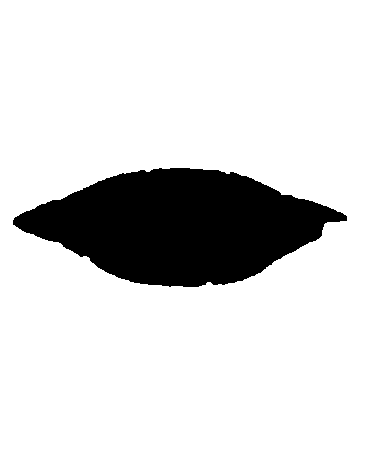}
    	\caption{Binary silhouette mask}
    	\label{fig:mask}
    \end{subfigure}
    \begin{subfigure}{.33\textwidth}
        \centering
        \includegraphics[width=0.5\linewidth]{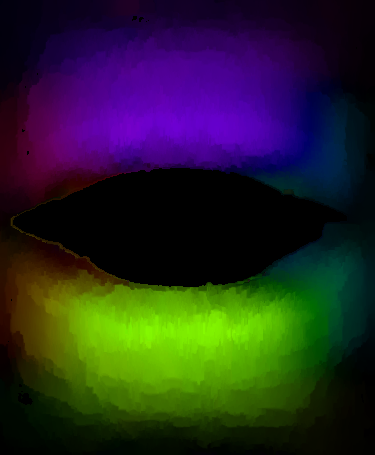}
    	\caption{ Applied silhouette mask to optical flow}
    	\label{fig:appmask}
    \end{subfigure}
    \caption{
    (a) OpenCV's \cite{opencv_library} sparse to dense optical flow algorithm interpolates a flow in the gap,
    (b) shows the binary silhouette mask that identifies the gap,
    (c) shows the cleaned optical flow with velocities set to zero in the gap.
    }
    \label{fig:ofs}
\end{figure}

\paragraph{High-speed Camera}
The high-speed camera provides a sequence of mono-chromatic images, as shown in Fig.~\ref{fig:vfhsc}. 
These images can be described as a function of intensities $I(\vy,t_i) : \mathbb{R}^2 \times \mathbb{N}_0 \rightarrow \mathbb{R}$ with an image resolution $w \times h$, the screen-space coordinates $\vy \in [0,w-1]\times[0,h-1]\subset\mathbb{R}^2$ and with $t_i \in [0,F) \subset \mathbb{N}_0$ being the currently sampled frame of the video with $F$ many frames in total.
Then, $u_I(\vy,t)$ and $v_I(\vy,t)$ denote the estimated motion of the screen-space coordinate $\vy$ at time $t$ along the longitudinal and lateral axis, respectively, resulting in the motion vector field:
\begin{align}
    \vv_I(\vy,t) = \begin{pmatrix}
    u_I(\vy,t) \\
    v_I(\vy,t)
    \end{pmatrix}.
    \label{eq:velocity-horizontal}
\end{align}
We applied the sparse to dense optical flow algorithm based on the Lucas-Kanade feature tracker with pyramids by \cite{lkpyr}, which is available in OpenCV~\cite{opencv_library}.
Such dense optical flow algorithm estimates a motion for every pixel of the input image, but tends to oversmooth flows (see Fig.~\ref{fig:osof}), which produces artifacts at edges.
To mitigate this problem, we apply a mask (see Fig.~\ref{fig:mask}), which helps to distinguish the vocal folds from the gap between them.
We use the intensity values for the mask segmentation with a subsequent median filter for more robustness to noise. 
Using the mask (see Fig.~\ref{fig:appmask}), we set the motion vectors inside the gap explicitly to zero vectors, since there is no vocal fold movement. 
For consistency with the remaining data, the motion in the image plane is converted to physical domain units. 
The high-speed images and the resulting optical flows are discrete in space (individual pixels) and in time (individual frames).
We interpolate the resulting discrete vector piecewise bilinearly in space with $C^0$ continuity and we interpolate quadratically with $C^1$ continuity in time to extract a continuous time-dependent two-dimensional vector field.

\begin{figure}[b]
	\centering
	\includegraphics[width=0.1667\linewidth]{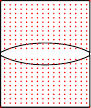}
	\caption{
        Schematic view of the vibrometer measurement points marked in red on the vocal fold model.
    }
	\label{fig:vmmp}
\end{figure}

\paragraph{Vibrometer Measurements}
Vibrometer measurements are taken at discrete grid points (see Fig.~\ref{fig:vmmp}) with screen-space coordinates $\vy_i$, which remain the same across all time steps.
For each measurement point, we receive an associated velocity component $w_i(\vy_i,t_i)$ along the vertical axis.
Unlike the optical flow vector fields, the measurement points cannot be interpolated bilinearly in space, because the measurement grid is not exactly regular.
Instead, we interpolate spatially using radial basis functions:
\begin{align}
    w(\vy,t_i) = \sum_i \omega_i \cdot \phi(\|\vy - \vy_i\|)
    \label{eq:velocity-vertical}
\end{align}
where $\phi(r) = r^2 \ln(r)$ is the thin plate spline kernel and the weights $\omega_i$ are linearly fitted per time step $t_i$ to meet all constraints in a least-squares sense, i.e., $\sum_i\|w(\vy_i,t_i) - w_i(\vy_i,t_i)\|^2 \rightarrow \min$.
Formally, we interpolate the vibrometer measurements linearly in time to obtain a continuous field $w(\vy,t)$.
With this, $w(\vy,t)$ gives us the velocity for an arbitrary screen-space coordinate $\vy$ at time $t$.
Using the optical flow $\vv_I(\vy,t)$ from Eq.~\eqref{eq:velocity-horizontal}, we combine the longitudinal and lateral velocity components with the vertical component from the vibrometer in Eq.~\eqref{eq:velocity-vertical} to obtain the tri-variate surface vector field $\vv(\vy,t)$ that describes the observed three-dimensional motion of the vocal fold surface:
\begin{align}
    \vv(\vy,t) = \begin{pmatrix}
    u_I(\vy,t)\\
    v_I(\vy,t)\\
    \,w(\vy,t)
    \end{pmatrix}
    \label{eq:velocity-observed}
\end{align}
In the following, we optimize for a vocal fold model using FEM simulations that match the observed motion.

\subsection{Vocal Fold Model}

\begin{figure}[t]
	\centering
    \begin{subfigure}{.33\textwidth}
        \centering
        \includegraphics[width=0.5\linewidth]{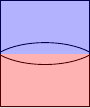}
    	\caption{Scalar fields for lateral component}
    	\label{fig:late}
    \end{subfigure}
    \begin{subfigure}{.33\textwidth}
        \centering
        \includegraphics[width=0.5\linewidth]{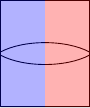}
    	\caption{Scalar fields for longitudinal component}
    	\label{fig:long}
    \end{subfigure}
    \begin{subfigure}{.33\textwidth}
        \centering
        \includegraphics[width=0.5\linewidth]{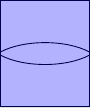}
    	\caption{Scalar field for vertical component}
    	\label{fig:vert}
    \end{subfigure}
	\caption{
    To model discontinuities, the force field is split in the lateral and longitudinal direction into two pieces with $C^0$ continuity. a) shows the two lateral pieces, b) shows the two longitudinal pieces, and c) shows that there is only one piece used for the vertical component. The same layout is used on the superior and inferior side of the vocal folds.}
	\label{fig:sepa}
\end{figure}

We parameterized the inferior and superior surface of the M5 vocal fold model with $(u,v)$ coordinates.
Thus, given is the M5 vocal fold model with a surface $\vx_0(u,v) : \mathbb{R}^2 \rightarrow \mathbb{R}^3$, parameterized by $(u,v)$ coordinates.
For each time $t$, the model is subject to a stationary FEM simulation, resulting in a deformed shape $\overline\vx(u,v,t)$.
Let $\vf(u,v,t)$ be the unknown time-dependent force field acting on the surface. 
We compose the force field from a tensor-product with unknown control forces $\vf_{i,j,k}$. 
Thereby, the polynomial degrees are $l$, $m$, and $n$, and the B\'ezier Bernstein basis functions are $B_i^n(t) = \binom{n}{i} \; t^i \; (1-t)^{n-i}$.
A force field $\vf(u,v,t)$ is given as tensor product via:
\begin{align}
\vf(u,v,t) = \sum_{i=0}^l \sum_{j=0}^m \sum_{k=0}^n \vf_{i,j,k} \; B_i^l(u) \; B_j^m(v) \; B_k^n(t)
\end{align}
Discontinuities in the force field are required due to discontinuities in the observable motion.
For example, a discontinuity in the velocity along the lateral axis can be observed when the vocal folds touch and are pushed apart by the air flow.
Thus, we separate our force field into 10 individual scalar fields to have greater control over each individual component and to model discontinuities in our force field.
We employ five of them to act on the inferior surface and five of them on the superior surface of the vocal folds.
Two of the five scalar fields represent the longitudinal component of the force, two scalar fields represent the lateral component of the force and one scalar field represents the vertical component of the force.
The scalar fields for the longitudinal and lateral axis cover only half of the vocal folds as shown in Fig.~\ref{fig:sepa}.

\begin{figure}
    \centering
    \begin{subfigure}{.45\textwidth}
        \centering
        \includegraphics[width=1\linewidth]{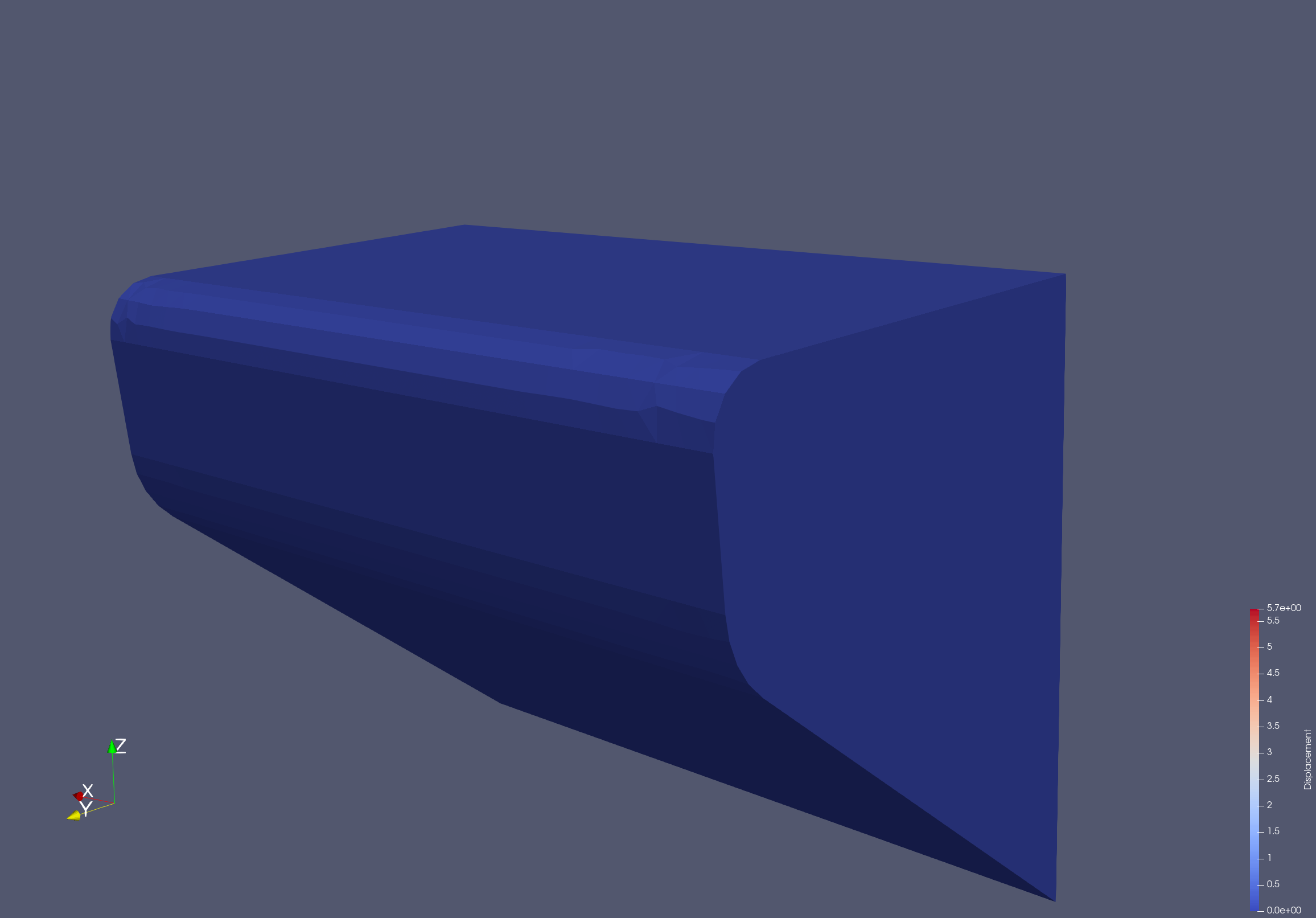}
    	\caption[Vocal fold in rest]{Rest state (before applying forces)}
    	\label{fig:vfre}
    \end{subfigure}
    \quad%
    \begin{subfigure}{.45\textwidth}
        \centering
        \includegraphics[width=1\linewidth]{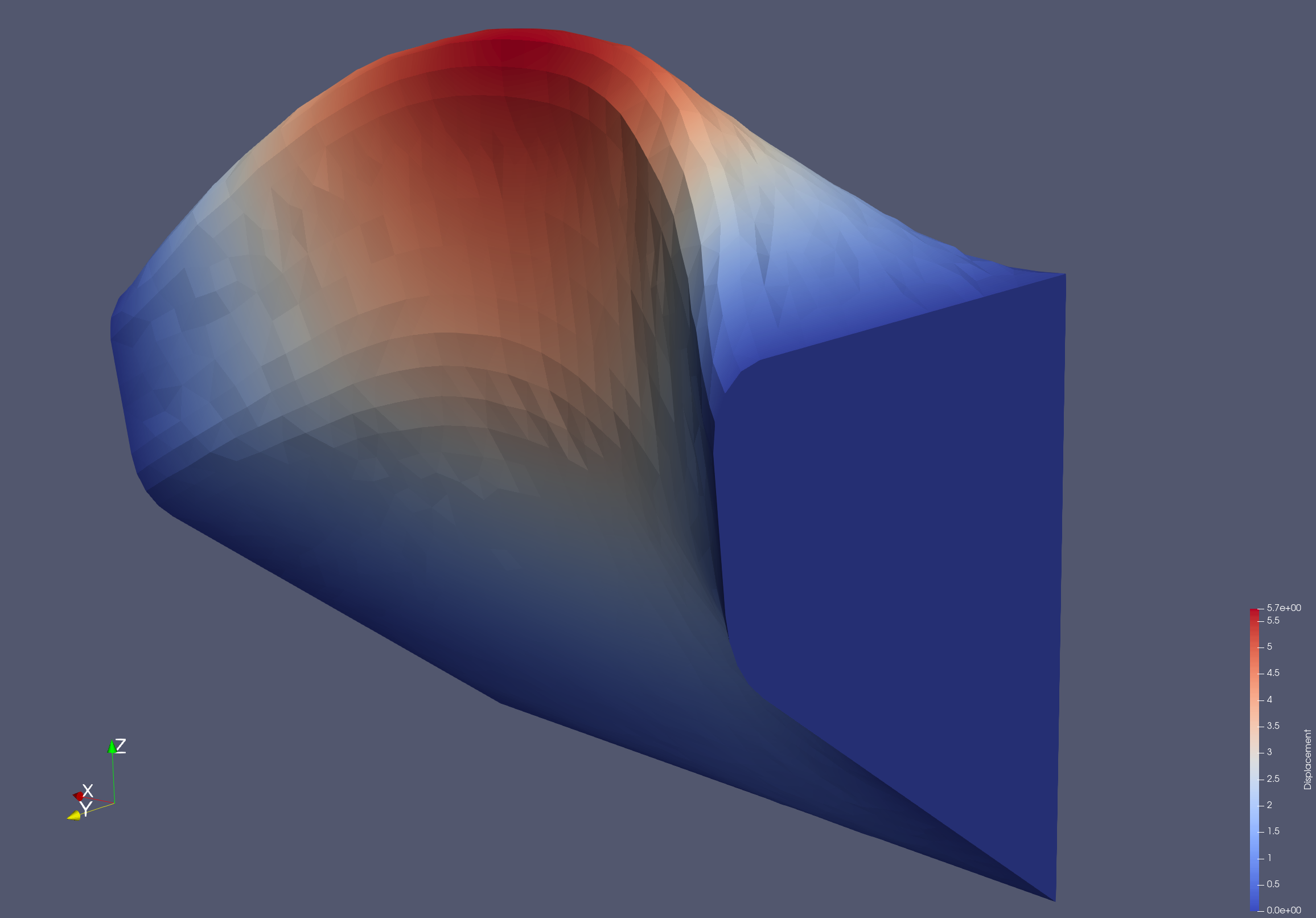}
    	\caption[Vocal fold in equilibrium]{Equilibrium state (after applying forces)}
    	\label{fig:vfeq}
    \end{subfigure}
    \caption{Comparison of one vocal fold before and after the FEM simulation, where the displacement magnitude is mapped to color. (a) shows the initial rest position without forces applied to it, and (b) shows the vocal fold in its equilibrium state after the forces were applied.
    }
    \label{fig:vfreeq}
\end{figure}

\paragraph{Finite Element Method}
We utilize FEM simulations for the deformation of the initial vocal fold model.
We discretize the vocal fold model and apply the sampled forces from the force field function $\vf$ to the inferior and superior areas of the model. 
The FEM simulation requires Lam\'e constants $\lambda$ and $\mu$, which determine the stretching and stiffness of the material. 
$\lambda$ and $\mu$ can be calculated using the Poisson ratio $\nu$ and the Young's modulus $E$ of the used elastomer:
\begin{align}
    \lambda&=\frac{E\nu}{\left(1+\nu\right)\left(1-2\nu\right)}
    &
    \mu&=\frac{E}{2\left(1+\nu\right)}
\end{align}
We define the stress $\sigma(\vx)  : \mathbb{R}^3 \rightarrow \mathbb{R}^{3 \times 3}$ as tensor field for position $\vx$ and denote the resulting displacement vector field as $\vu(\vx) : \mathbb{R}^3 \rightarrow \mathbb{R}^3$.
The FEM simulation solves the linear elasticity partial differential equation after the forces are applied with $\mI$ being the identity matrix and $\vf(\vx) : \mathbb{R}^3 \rightarrow \mathbb{R}^3$ being the externally applied forces:
\begin{align}
    \vf(\vx) + \left( \nabla \cdot \sigma\left( \vx \right)^{\textrm{T}} \right)^{\textrm{T}} &=0\\
    \sigma\left(\vx\right)&=\lambda\;\textrm{div}\left(\vu(\vx)\right)\mI+\mu\left(\nabla \vu(\vx)+\nabla \vu(\vx)^{\textrm{T}}\right)
\end{align}
The resulting displacement field creates a deformation of the vocal folds (see Fig.~\ref{fig:vfeq}), where the externally applied forces are in an equilibrium with the internal forces of the deformed material.
We additionally set boundary conditions on the sides that are in contact with the fixture.
These sides are immovable, because the sides of the model are glued to the fixture in the experimental setup and therefore cannot move.

\paragraph{Vocal Fold Motion}
Our goal is to match the motion of the simulated vocal fold with the motion observed in the measurement data.
Next, we describe how the motion of the simulated vocal fold is calculated in screen-space, which is the space in which simulation and measurement are compared.
For the screen-space coordinates $\vy$, we determine the $(u,v)$ coordinates of the visible surface of the deformed model $\overline\vx(u,v,t)$ by ray casting.
Thereby, the binary indicator $m(\vy) \in \{0,1\}$ is created, which denotes whether the ray cast was successful, i.e., whether a vocal fold surface is visible in the pixel at screen-space coordinate $\vy$.
The screen-space motion is then determined by finite differences:
\begin{align}
    \overline\vv(\vy,t) := \frac{\diff \overline\vx}{\diff t}(u,v,t) \approx \frac{
    \overline \vx(u,v,t+\Delta t) - \overline \vx(u,v,t)}{\Delta t}
    \label{eq:velocity-simulated}
\end{align}
where the mesh points with the same $(u,v)$ coordinates are taken, namely the coordinates observed at time $t$ for pixel $\vy$.
With this, we observe where a mesh point with a certain $(u,v)$ coordinate moves to over time in screen-space.

\begin{figure}[ht]
	\centering
    \includegraphics[width=0.17\linewidth]{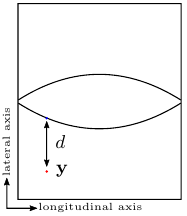}%
    \vspace{-0.5em}%
	\caption{
    The gap opening function $o(\vy,t)$ returns for a given screen-space coordinate $\vy$ (red) the unsigned distance $d$ to the closest point on the silhouette (blue) along the lateral axis.}
	\label{fig:ofunc}
\end{figure}

\paragraph{Gap}
The gap is the opening between the two individual vocal fold halves.
The gap is a defining feature of the vocal fold motion and we thus aim to preserve the gap in our FEM-based reconstruction.
To measure the extent of the gap, we introduce the gap opening function $o(\vy,t) : \mathbb{R}^2 \times \mathbb{R} \rightarrow \mathbb{R}^+$, from which the silhouette of the gap can be determined, as illustrated in Fig.~\ref{fig:ofunc}.
For a given screen-space coordinate $\vy$ at time $t$ the gap opening function $o(\vy,t)$ denotes the unsigned lateral distance to the closest position on the gap silhouette, which is determined by casting a ray.
If the ray does not intersect the gap, then the distance to the last known opening location is taken.

\subsection{Optimization}
\label{sec:opti}
In the following, we describe the energy minimization approach that determines the forces to apply in the FEM simulation.
We initialize the forces with zero and optimize them iteratively according to the error metric.

\paragraph{Error Metric}
We minimize an error $\mathrm{E}$, which penalizes a mismatch in the reproduced motion and a difference in the silhouette of the gap.
For the latter, we apply an arctangent to enhance the penalty on smaller differences. 
We integrate the error over a selected time domain, from $T_1$ to $T_2$. 
\begin{align}
    \mathrm{E} = \frac{1}{T_{2}-T_{1}}\int_{T_{1}}^{T_{2}}  \frac{1}{N} \sum_{i=1}^N \underbrace{ m(\vy_{i}) \; \| \vv\left(\vy_i,t\right) - \overline\vv(\vy_i,t)
    \|^2}_{\textrm{motion}}
    + \underbrace{\arctan \left( o(\vy_i,t) \right)}_{\textrm{gap}}  \;\diff t
\end{align}
The motion term of the error sums up the $L^2$ norm of the difference in the measured velocity $\vv$ from Eq.~\eqref{eq:velocity-observed} and the simulated velocity $\overline\vv$ from Eq.~\eqref{eq:velocity-simulated} over all measurement points, if a measurement point hits both the deformed model as well as the model in the experiment ($m(\vy_{i})=1$).
The measurement point is ignored if the deformed model and the ground truth measurement point do not have a valid velocity, i.e., when $m(\vy_{i})=0$.
If only the vocal folds in the experiment have a valid velocity for the measurement point, then the $L^2$ norm of the ground truth velocity is added.
The second term of the error metric aims to match the silhouette of the gap.
We sum up the unsigned distance returned from the gap opening function $o(\vy_i,t)$ after we adjusted their influence using the $\arctan$ function. 
Finally, the accumulated error is normalized by the number of measurement points $N$.

\paragraph{Gradient Descent}
The error metric $\mathrm{E}$ is evaluated in each optimization step for a given set of control forces $\vf_{i,j,k}$.
Finite differences are applied to each individual control point of the force field to construct the gradient of the error metric.
Afterwards, gradient descent is used to find the control forces that minimize the error:
\begin{align}
    \vf_{i,j,k}^{n+1} = \vf_{i,j,k}^n - \gamma \frac{\partial}{\partial \vf_{i,j,k}}\mathrm{E}(\vf_{i,j,k}^n)
\end{align}
A gradient descent with step size $\gamma$ is able to find a local minimum of a function by descending along the steepest descent direction.
To accelerate convergence, we perform a line search with increasing step size.

\begin{figure}[p]
    \centering
    \begin{minipage}{0.02\linewidth}
        ~
    \end{minipage}
    \begin{minipage}{0.97\linewidth}
        \centering
        \begin{minipage}{0.32\linewidth}
            \centering
            $3\times 3\times 10$ (one-sided forces)
            \makecircle{clrOneLow}
        \end{minipage}
        \begin{minipage}{0.32\linewidth}
            \centering
            $3\times 3\times 10$ (two-sided forces)
            \makecircle{clrTwoLow}
        \end{minipage}
        \begin{minipage}{0.32\linewidth}
            \centering
            $5\times 5\times 15$ (two-sided forces)
            \makecircle{clrTwoHigh}
        \end{minipage}
    \end{minipage}
    \centering
    \begin{minipage}{0.02\linewidth}
        $t_0$
    \end{minipage}
    \begin{minipage}{0.97\linewidth}
    \centering
    \includegraphics[width=0.32\linewidth]{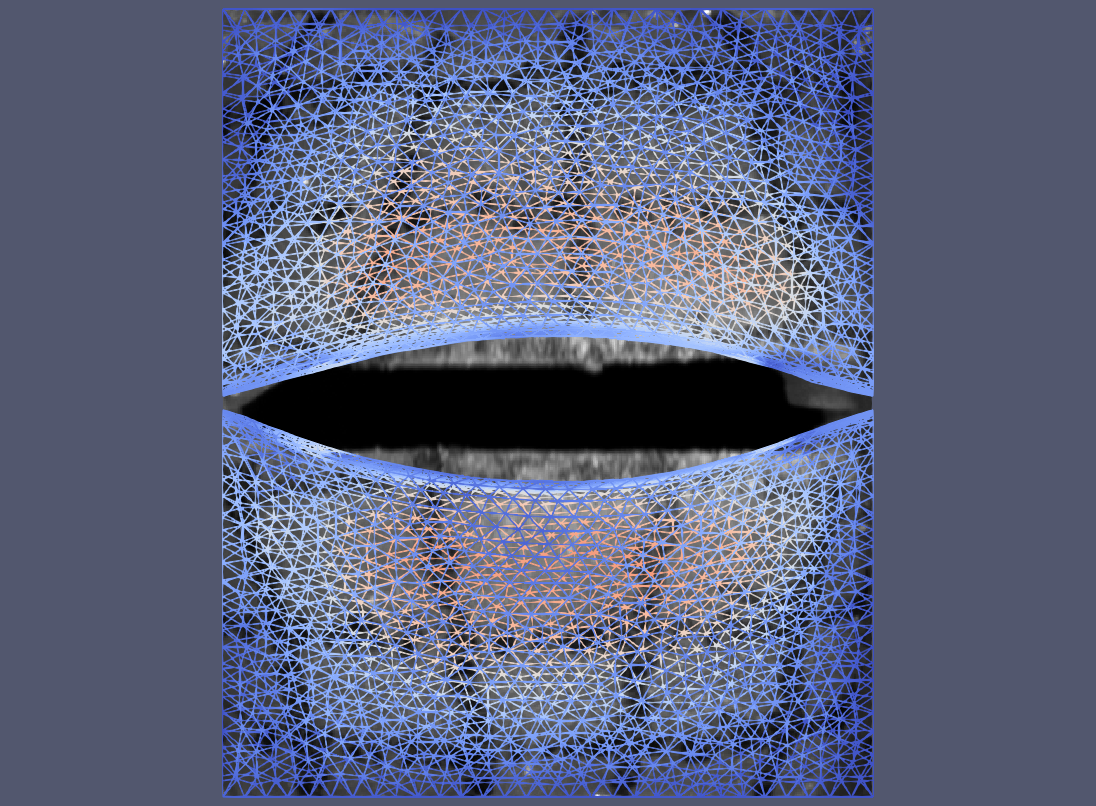}
    \includegraphics[width=0.32\linewidth]{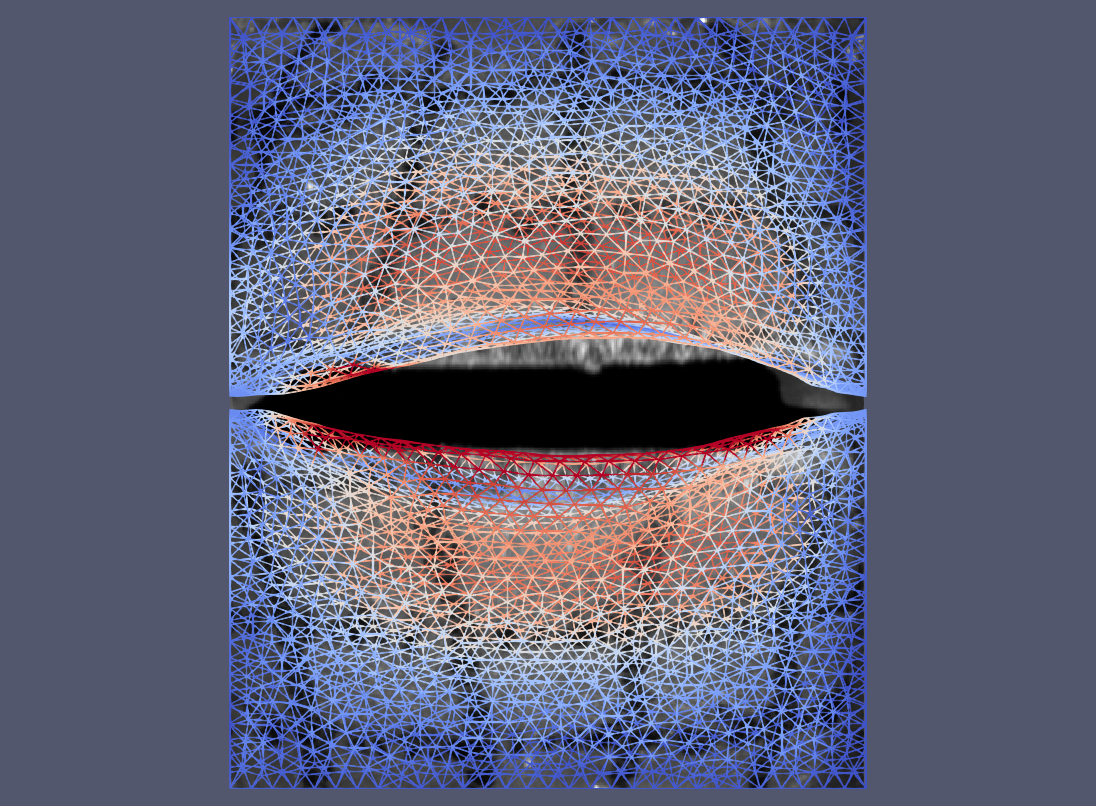}
    \includegraphics[width=0.32\linewidth]{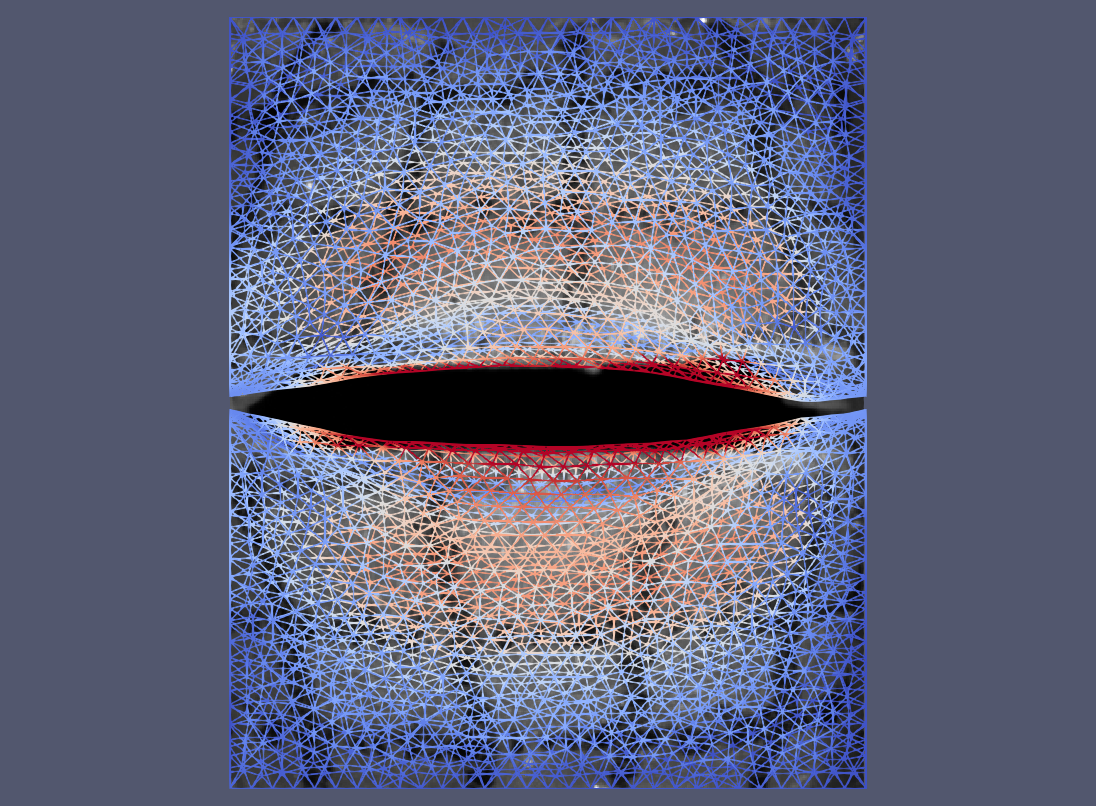}
    \end{minipage}
    \\
    \centering
    \begin{minipage}{0.02\linewidth}
        $t_{150}$
    \end{minipage}
    \begin{minipage}{0.97\linewidth}
    \centering
    \includegraphics[width=0.32\linewidth]{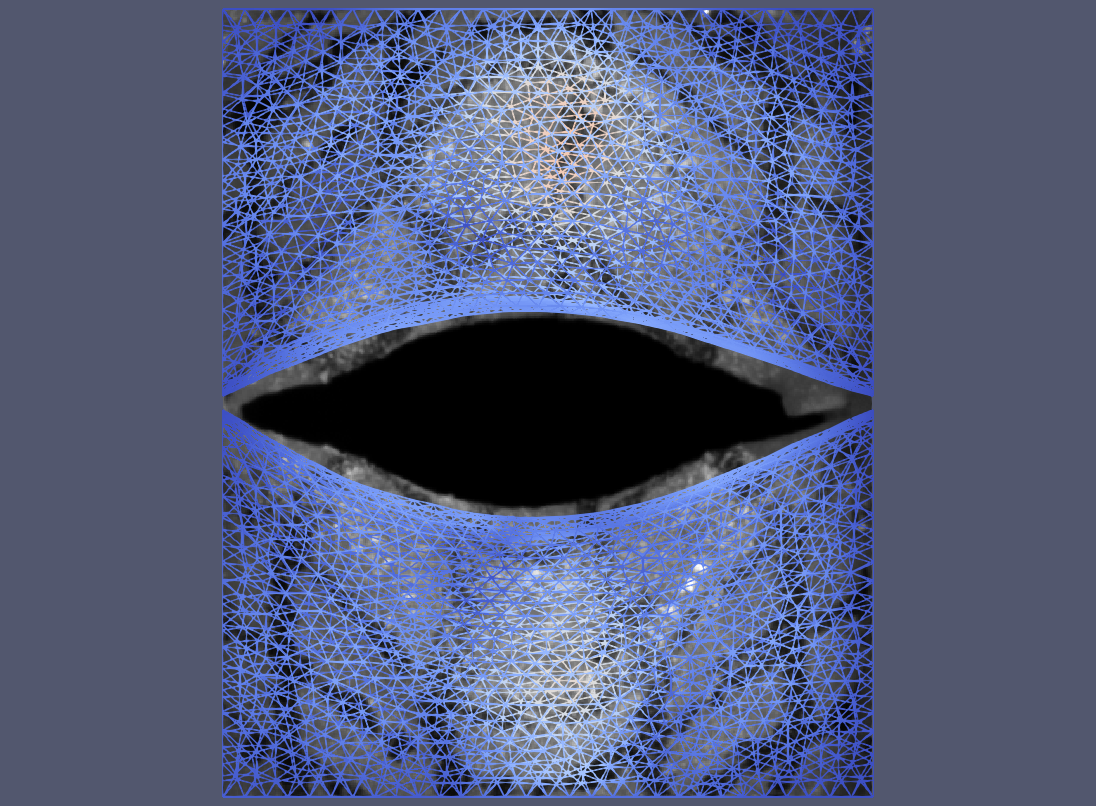}
    \includegraphics[width=0.32\linewidth]{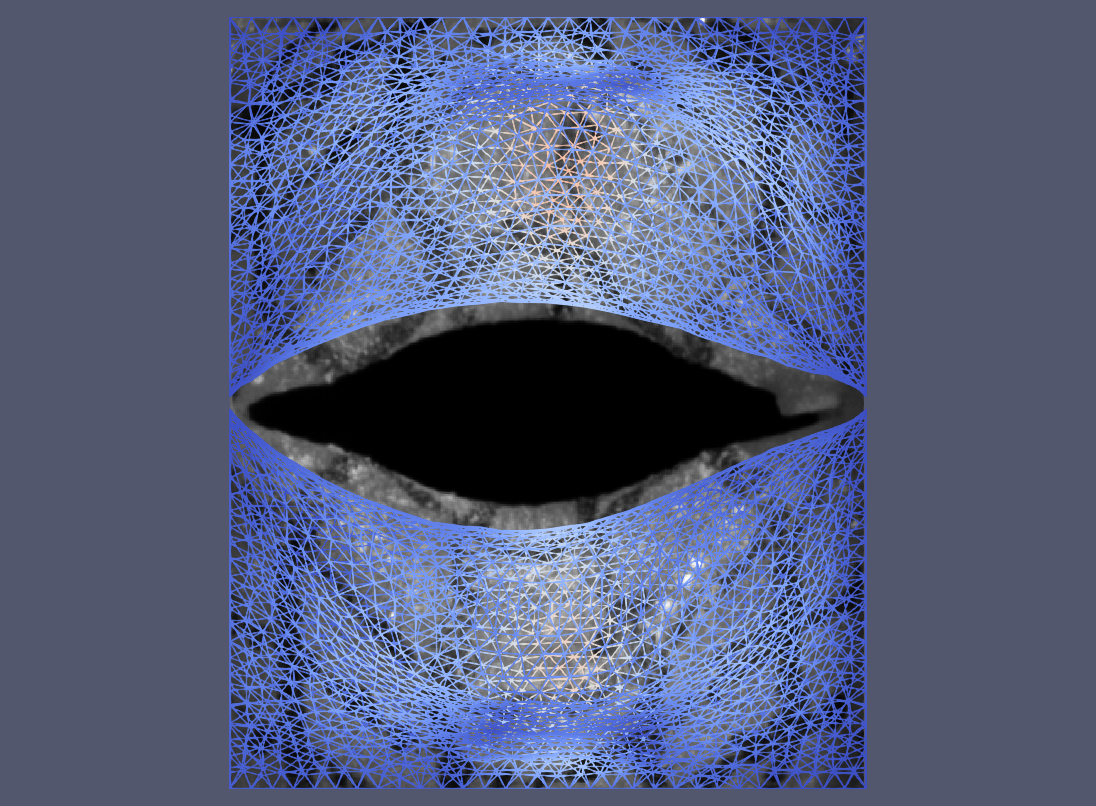}
    \includegraphics[width=0.32\linewidth]{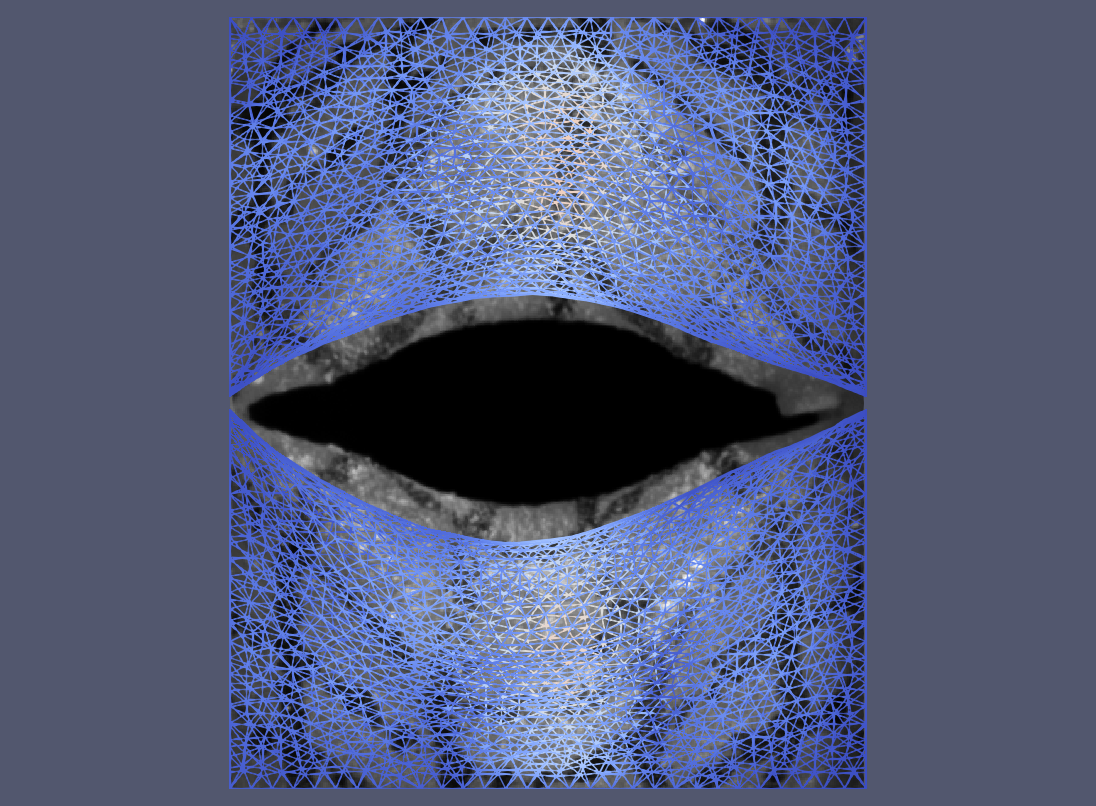}
    \end{minipage}
    \\
    \centering
    \begin{minipage}{0.02\linewidth}
        $t_{310}$
    \end{minipage}
    \begin{minipage}{0.97\linewidth}
    \centering
    \includegraphics[width=0.32\linewidth]{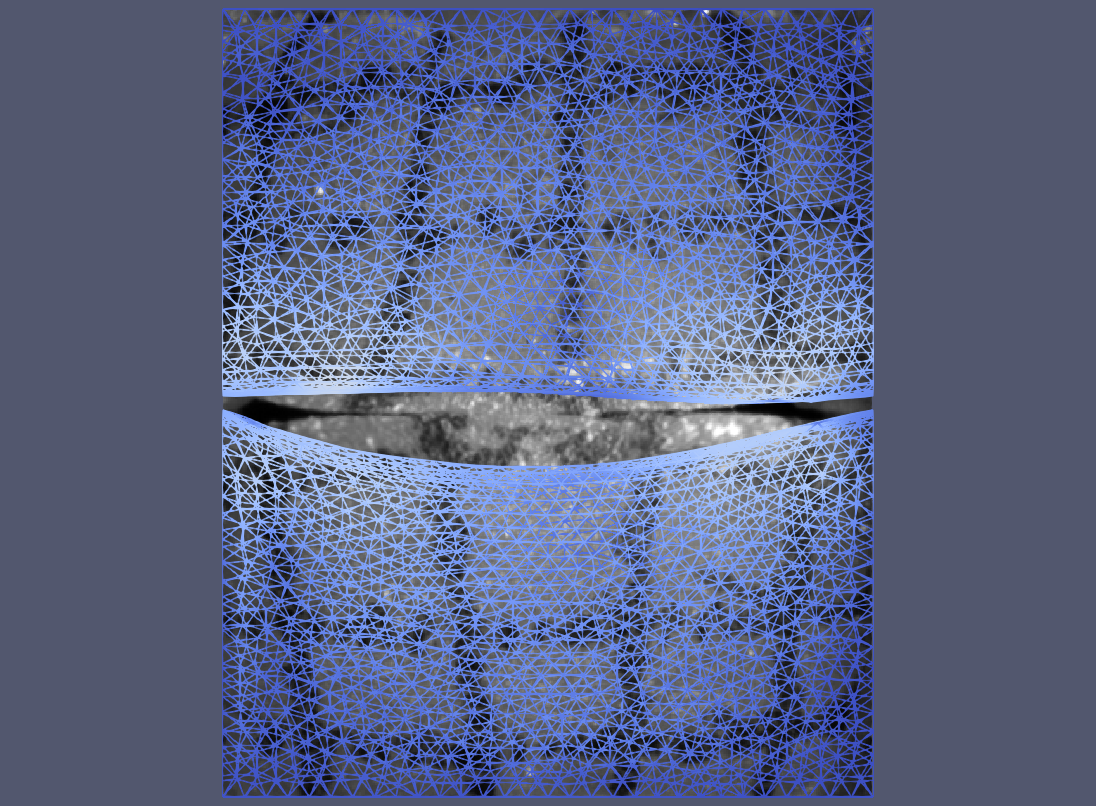}
    \includegraphics[width=0.32\linewidth]{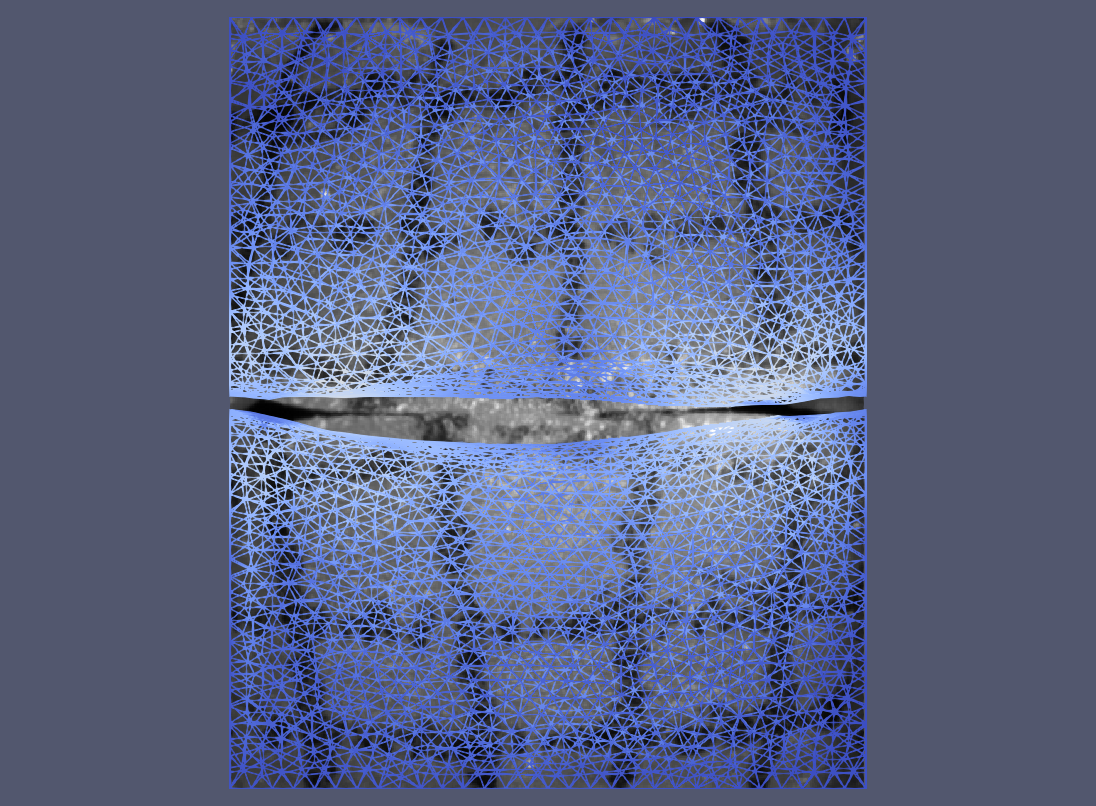}
    \includegraphics[width=0.32\linewidth]{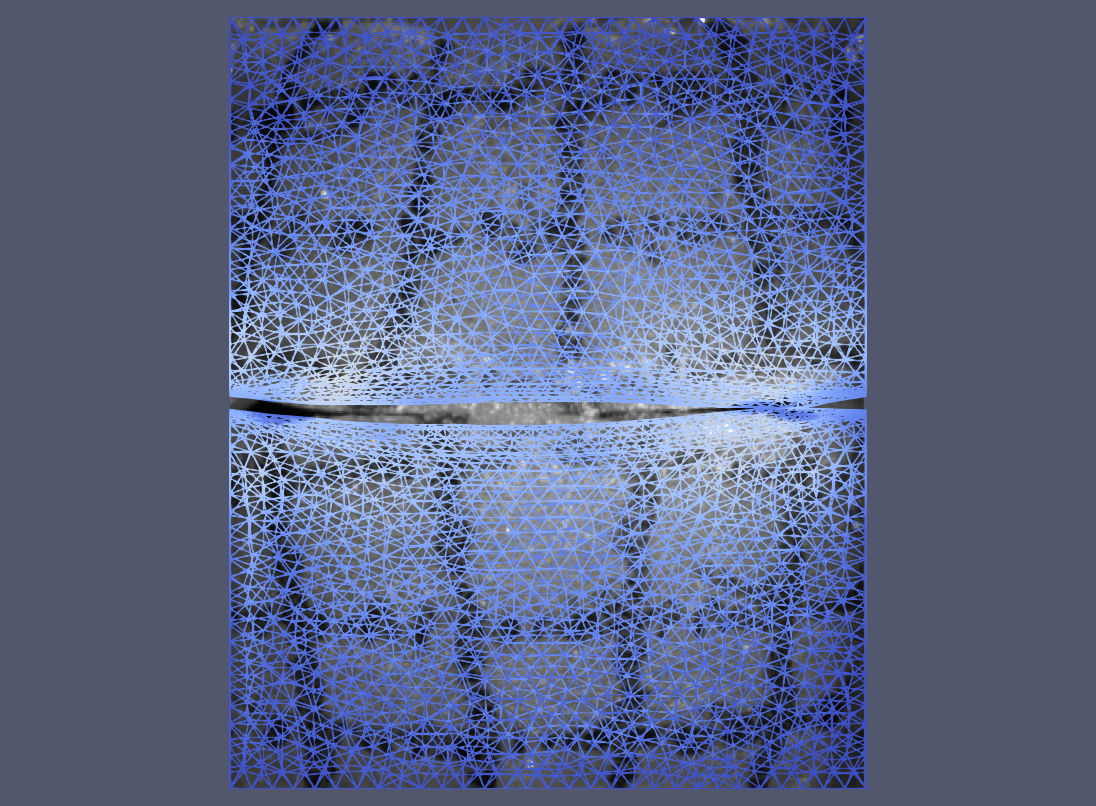}
    \end{minipage}
    \\
    \centering
    \begin{minipage}{0.02\linewidth}
        $t_{450}$
    \end{minipage}
    \begin{minipage}{0.97\linewidth}
    \centering
    \includegraphics[width=0.32\linewidth]{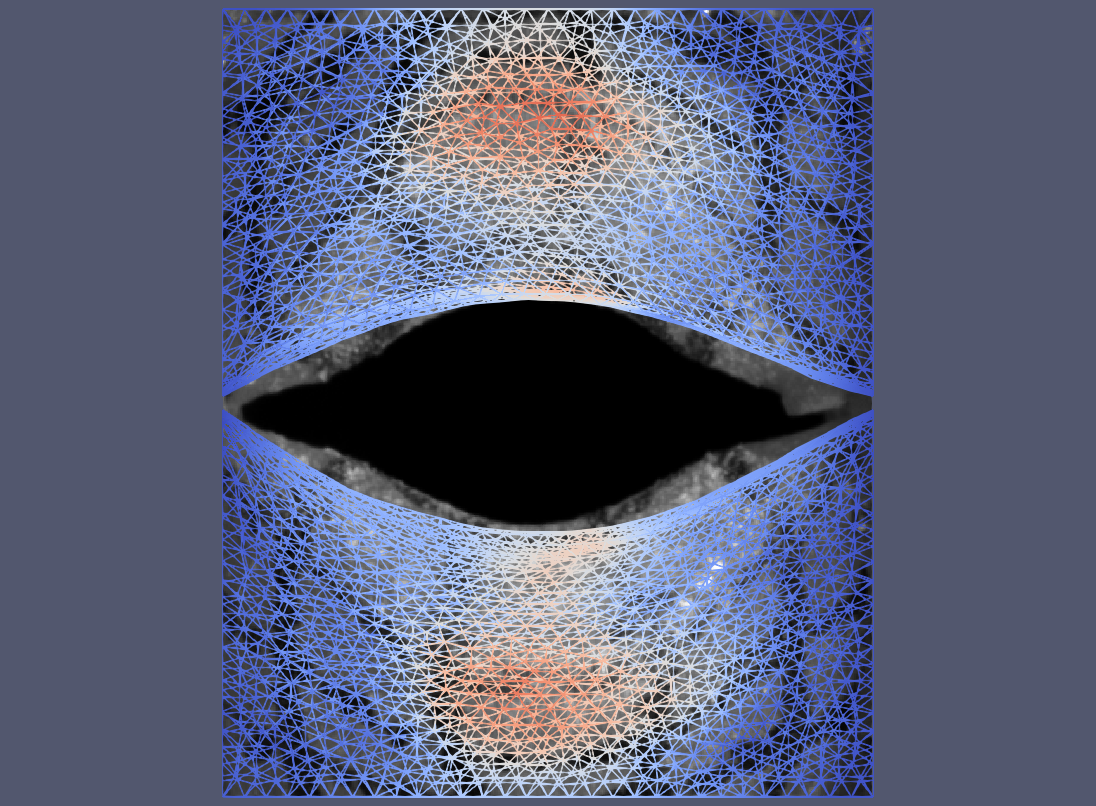}
    \includegraphics[width=0.32\linewidth]{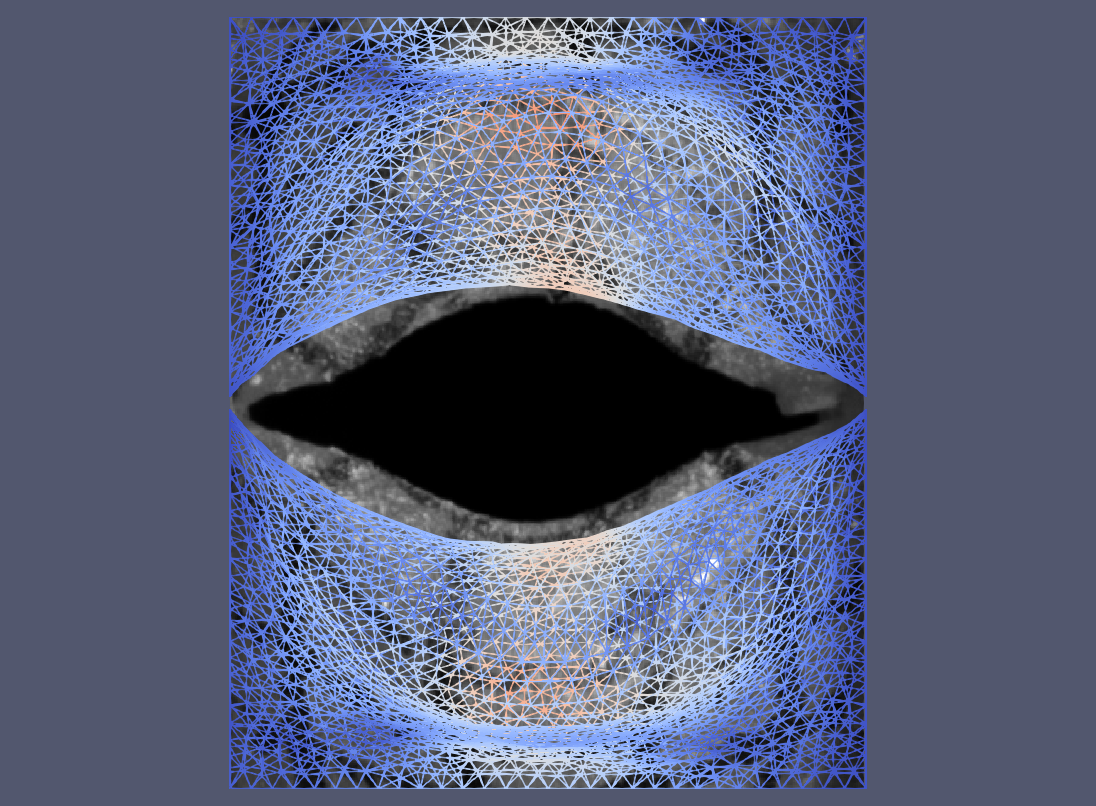}
    \includegraphics[width=0.32\linewidth]{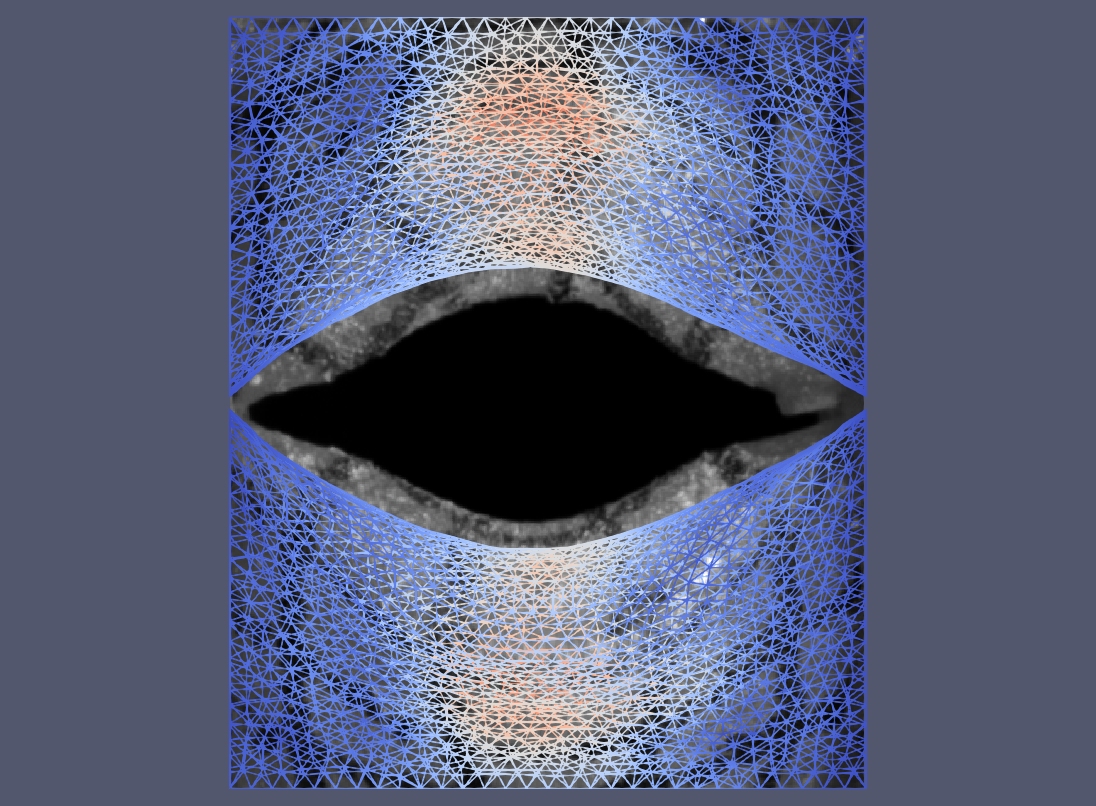}
    \end{minipage}
    \\
    \centering
    \begin{minipage}{0.02\linewidth}
        $t_{499}$
    \end{minipage}
    \begin{minipage}{0.97\linewidth}
    \centering
    \includegraphics[width=0.32\linewidth]{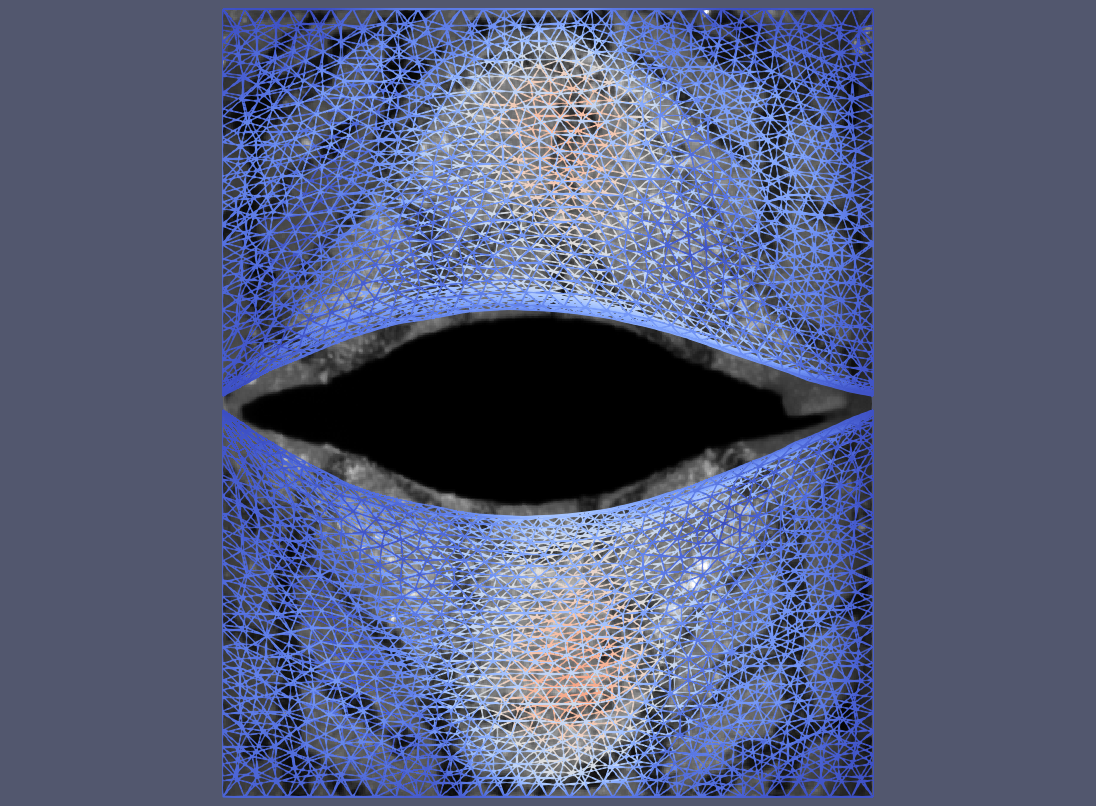}
    \includegraphics[width=0.32\linewidth]{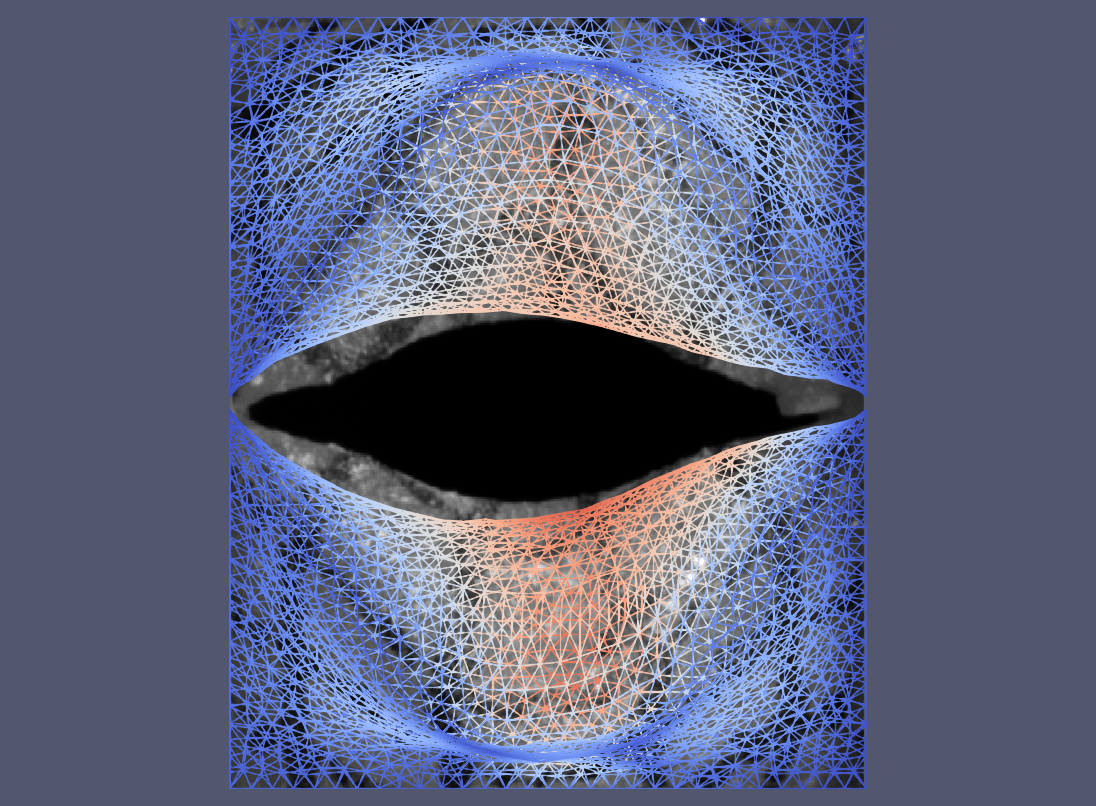}
    \includegraphics[width=0.32\linewidth]{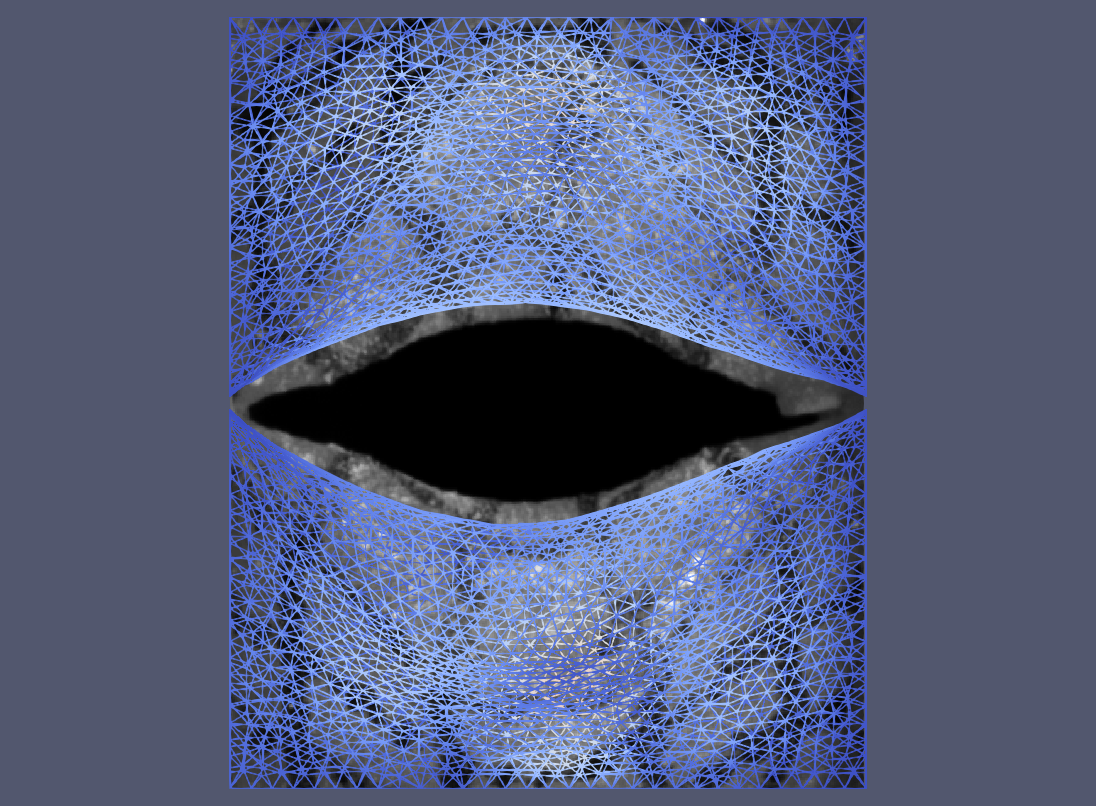}
    \end{minipage}
    \caption{Visual comparison of the three setups for different time steps of the animation. Left: optimization for a force field with $3 \times 3 \times 10$ control points (longitudinal, lateral, time) on the inferior side only. Middle: Adding a second force field on the superior side allows for a better capture of the bending of vocal folds. Right: Increasing the resolution to $5 \times 5 \times 15$ allows for bending while maintaining symmetry. The displacement magnitude is color-coded.}
    \label{fig:wireframe-animation}%
\end{figure}

\section{Results}
In the following section, we show the results of our vocal fold surface reconstruction, and we compare the motion of the FEM model with the velocity measurements.
The unknown in our optimization is the force field $\vf$ that is applied to the vocal fold surface in the FEM simulation.
To validate our choice of optimizing for a force field on both the inferior and the superior side of the vocal model, we performed two experiments: first, we applied the force field only on the inferior side, and second we applied the force field on both sides.
Another degree of freedom is the number of control points that are used to model the force field as a tensor product.
Here, we compared a $3\times 3\times 10$ resolution with a $5\times 5\times 15$ resolution (longitudinal, lateral, time).

\paragraph{Time-Series Comparison with Optical Measurements}
In Figure \ref{fig:wireframe-animation}, the optimized FEM models are overlayed on top of the optical high-speed camera measurements. 
The color encodes the displacement magnitude $\|\vu\|$.
The vocal fold shape obtained from the two-sided force field optimization matches the optical measurements better.
In particular, the opening of the gap and the folding of the vocal folds match the optical measurements more closely.
In the simulation results, however, an asymmetry in the vocal folds is apparent, which could indicate a pathology.
The asymmetry reduces with increasing tensor product resolution.
Closer inspection of the high-speed camera footage revealed that an asymmetry is also present in the experimental data, since the vocal folds were not glued in perfectly symmetric.

\begin{figure}[tp]
    \centering
    \includegraphics[width=0.45\linewidth]{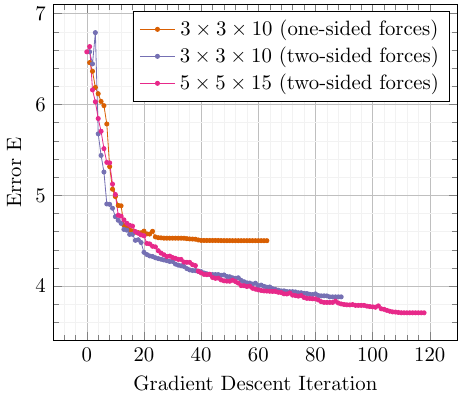}
    \caption{Convergence plots of error $\textrm{E}$ over the course of the gradient descent minimization. With one-sided forces, the error reaches a plateau after 25 gradient descent iterations. With two-sided forces, the error can be reduced further.}
    \label{fig:convergence-plot}
\end{figure}

\paragraph{Convergence}
The error convergence plots of the gradient descent schemes are shown for all experiments in Figure~\ref{fig:convergence-plot}, where the optimization was terminated once the error minimization converged.
While the one-sided force field optimization converged after about 25 iterations, the two-sided force field optimization converged after 85 iterations and 115 iterations, respectively. 
Both two-sided force field optimizations had a similar error decay.
The model with higher resolution, however, was able to reduce the error a bit further before reaching convergence.

\begin{figure}[p]
\centering
\begin{minipage}{0.05\linewidth}
    \centering
    ~
\end{minipage}
\begin{minipage}{0.48\linewidth}
    \centering
    $\vy_i = (7.45, 4.54)$
\end{minipage}
\begin{minipage}{0.45\linewidth}
    \centering
    $\vy_i = (12.73, 15.25)$
\end{minipage}
\\%
\begin{subfigure}{.49\textwidth}
\includegraphics[height=0.75\linewidth]{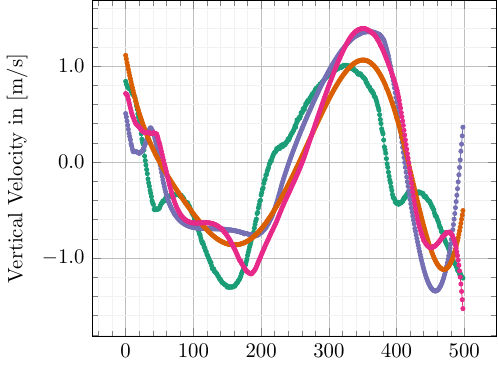}
\end{subfigure}%
\hfill%
\begin{subfigure}{.49\textwidth}
\includegraphics[height=0.75\linewidth]{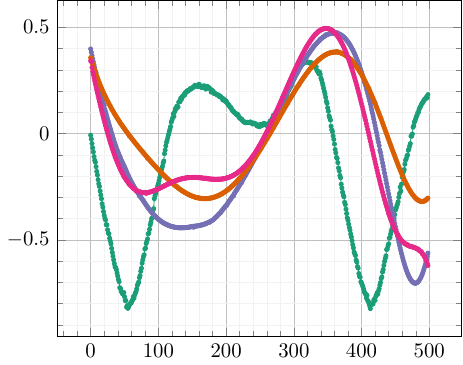}
\end{subfigure}
\\%
\begin{subfigure}{.49\linewidth}
\includegraphics[height=0.75\linewidth]{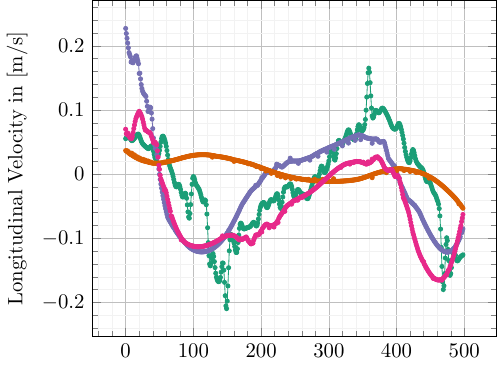}
\end{subfigure}%
\hfill%
\begin{subfigure}{.49\linewidth}
\includegraphics[height=0.75\linewidth]{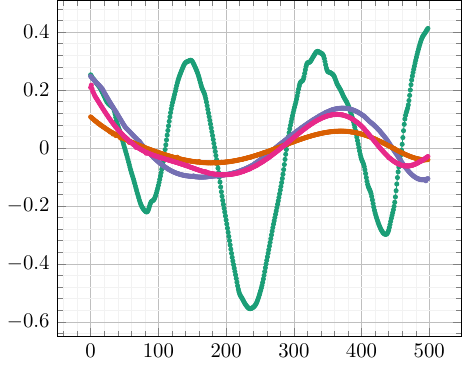}
\end{subfigure}
\\%
\begin{subfigure}{.49\linewidth}
\includegraphics[height=0.817\linewidth]{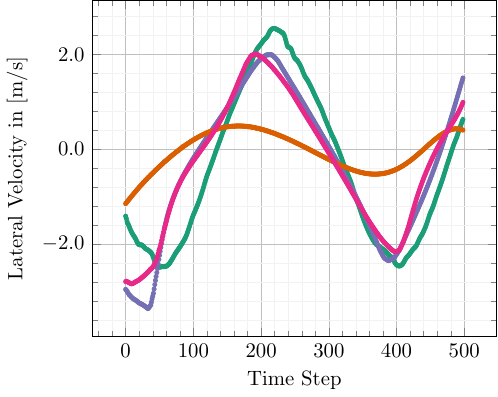}
\end{subfigure}%
\hfill%
\begin{subfigure}{.49\linewidth}
\includegraphics[height=0.827\linewidth]{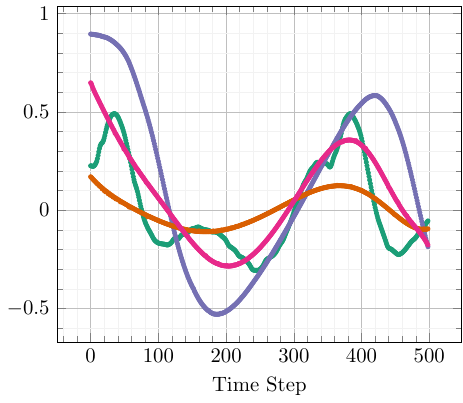}
\end{subfigure}%
\caption{Plots of vertical, longitudinal, and lateral velocity components (in rows) for 
vibrometer measurements~\makecircleem{clrVibr}, our
$3\times 3\times 10$ (one-sided forces) \makecircleem{clrOneLow},
$3\times 3\times 10$ (two-sided forces) \makecircleem{clrTwoLow}, and
$5\times 5\times 15$ (two-sided forces) \makecircleem{clrTwoHigh} for two different selected image coordinates (in columns). An increase in force field resolution and a two-sided force field application result in the lowest error. Residuals are visible in particular for high-frequent temporal changes due to insufficient degrees of freedom in the optimization.}
\label{fig:time-plot-pnts}
\end{figure}

\paragraph{Agreement with Observed Velocities for Selected Surface Points}
Aside from matching the silhouette of the gap, the objective function aims to reconstruct the observed velocities on the superior side, considering both the vibrometer measurement (vertical component) and the optical flow (horizontal components).
In Figure~\ref{fig:time-plot-pnts}, we show the velocity signal over time for two selected points on the vocal fold surface.
The velocity signals are shown for the vertical, longitudinal, and lateral component, separately.
For the first observation point (left column), the two-sided force fields were closer to the measured data than the one-sided model.
An increase in resolution of the force field tensor product resulted in marginally lower errors, since the fold had more degrees of freedom to match high-frequency fluctuations.
Note, however, that an exact reproduction of all high-frequent details is not desirable, since some can be considered measurement noise.
The second observation point (right column) exhibited for the longitudinal component a temporal variation that could not be matched by any of the three optimizations, due to an insufficient number of degrees of freedom.
The two-sided high-resolution optimization reproduced the lateral component noticeably better, whereas the differences in the vertical components were marginally improved.

\section{Conclusion}

In this paper, we proposed a novel FEM-based approach to reconstruct the three-dimensional motion of a complete vocal fold geometry based on optical measurements that were taken from the superior side only.
First, we reconstructed a surface velocity field from the optical flow of a high-speed camera recording (horizontal components) and a laser Doppler vibrometer (vertical component).
We then applied a time-dependent force field in tensor product basis to the inferior and/or superior side of the vocal model and let an FEM simulation deform the surface into an equilibrium state.
The force field control points are then optimized via gradient descent to minimize an error metric that consisted of two terms.
The first term measures the difference between simulated velocities and observed measurements, while the second term measures the silhouette difference between observation and simulation.
The results showed that our approach is able to qualitatively reproduce the underlying ground truth data of surface velocity and glottal gap. Increasing the resolution of the model resulted in a further reduction of the error. We saw that for the surface velocities, our models were able to reproduce the measurement data roughly. They were, however, not able to reproduce high-frequency components of the data. This has on the one hand the advantage, that it filters out measurement noise; on the other hand, this might also cause an unwanted low-pass filtering of the physical vocal fold surface velocities.

In future works, the behavior of our model in the frequency domain could be analyzed by optimization over several oscillation periods, followed by a transformation of the velocity data into the frequency domain. Furthermore, an increased accuracy might be achieved by a further increase of the model resolution. In both cases, an improved computation of the gradient needed for optimization like automatic differentiation would benefit the procedure by enhancing the computational efficiency of the numerical optimization, rendering more complex models possible. As another step, also dynamic effects of the vocal folds caused by e.g. viscoelasticity could be incorporated. 

\subsection*{Acknowledgements}
This work is funded by the German Research Foundation (DFG) through the projects ”Tracing the mechanisms that generate tonal content in voiced speech” (project number: 446965891) and "End-to-End Optimization for Energy-Driven Scientific Visualization" (project number: 517157369).

\end{document}